\documentclass[a4paper,twocolumn,longbibliography,accepted=2022-01-27]{quantumarticle}
\pdfoutput=1

\usepackage{amsmath,braket,amssymb}
\usepackage[final]{graphicx}
\usepackage{subfigure}
\usepackage{xcolor}
\usepackage[english]{babel}
\usepackage[utf8]{inputenc}
\usepackage{color}

\usepackage[sort&compress,numbers]{natbib}
\usepackage{hyperref}

\usepackage[normalem]{ulem}

\definecolor{forestgreen}{rgb}{0.08, 0.4, 0.13}
\definecolor{darkBlue}{rgb}{0.08, 0.13, 0.4}
\definecolor{changeRed}{rgb}{1, 0.2, 0.2}

\newcommand{\ave}[1]{\langle{#1}\rangle}

\makeatletter
\renewcommand{\fnum@figure}{FIG. \thefigure}
\makeatother

\begin{document}

\title{Dissipative Floquet Dynamics: \!\!\!\!\!\!\!\!\!\!\!\!\!\!\!
from Steady State to Measurement Induced Criticality in Trapped-ion Chains}

\author{Piotr Sierant}
\affiliation{The Abdus Salam International Center for Theoretical Physics, Strada Costiera 11, 34151 Trieste, Italy}
 \affiliation{Institute of Theoretical Physics, Jagiellonian University in Krakow, \L{}ojasiewicza 11, 30-348 Krak\'ow, Poland }
\author{Giuliano Chiriacò}
\affiliation{The Abdus Salam International Center for Theoretical Physics, Strada Costiera 11, 34151 Trieste, Italy}
\affiliation{SISSA — International School of Advanced Studies, via Bonomea 265, 34136 Trieste, Italy}
\author{Federica M. Surace}
\affiliation{The Abdus Salam International Center for Theoretical Physics, Strada Costiera 11, 34151 Trieste, Italy}
\affiliation{SISSA — International School of Advanced Studies, via Bonomea 265, 34136 Trieste, Italy}
\author{Shraddha Sharma}
\affiliation{The Abdus Salam International Center for Theoretical Physics, Strada Costiera 11, 34151 Trieste, Italy}
\author{Xhek Turkeshi}
\affiliation{The Abdus Salam International Center for Theoretical Physics, Strada Costiera 11, 34151 Trieste, Italy}
\affiliation{SISSA — International School of Advanced Studies, via Bonomea 265, 34136 Trieste, Italy}
\author{Marcello Dalmonte}
\affiliation{The Abdus Salam International Center for Theoretical Physics, Strada Costiera 11, 34151 Trieste, Italy}
\affiliation{SISSA — International School of Advanced Studies, via Bonomea 265, 34136 Trieste, Italy}
\author{Rosario Fazio}
\affiliation{The Abdus Salam International Center for Theoretical Physics, Strada Costiera 11, 34151 Trieste, Italy}
\affiliation{Dipartimento di Fisica, Universit\`a di Napoli ``Federico II'', Monte S. Angelo, I-80126 Napoli, Italy}
\author{Guido Pagano}
\email{pagano@rice.edu}
\affiliation{Department of Physics and Astronomy, Rice University, 6100 Main Street, Houston, TX 77005, USA}


\begin{abstract}
Quantum systems evolving unitarily and subject to quantum measurements exhibit various types of non-equilibrium phase transitions, arising from the competition between unitary evolution and measurements. Dissipative phase transitions in steady states of time-independent Liouvillians and measurement induced phase transitions at the level of quantum trajectories are two primary examples of such transitions. Investigating a many-body spin system subject to periodic resetting measurements, we argue that many-body dissipative Floquet dynamics provides a natural framework to analyze both types of transitions. We show that a dissipative phase transition between a ferromagnetic ordered phase and a paramagnetic disordered phase emerges for long-range systems as a function of measurement probabilities. A measurement induced transition of the entanglement entropy between volume law scaling and sub-volume law scaling is also present, and is distinct from the ordering transition. The two phases correspond to an error-correcting and a quantum-Zeno regimes, respectively. The ferromagnetic phase is lost for short range interactions, while the volume law phase of the entanglement is enhanced. An analysis of multifractal properties of wave function in Hilbert space provides a common perspective on both types of transitions in the system. Our findings are immediately relevant to trapped ion experiments, for which we detail a blueprint proposal based on currently available platforms.
\end{abstract}

\maketitle

\section{Introduction}\label{Sec:Intro}

The interplay of coherent and incoherent dynamics has a rich and long history in the context of quantum physics. At the basic quantum mechanical level, it is responsible for a plethora of few body effects, of particular relevance to quantum optical systems. Over the last two decades, boosted by impressive progress in both solid state and atomic experiments, this interplay has found vast application in the context of many-body phenomena as well.

A particularly successful framework in this respect has been the identification of new phases of matter associated with the non-equilibrium steady states of local Liouville dynamics. Such phases, understood as steady states of time-independent Liouvillians, have highly non-thermal properties and have been shown to often lack an equilibrium counterpart \cite{Sieberer16,Lee13,Jin16,Maghrebi2016,Overbeck2017,FossFeig2017,Jin18,Fink17,Fitzpatrick17,Flaschner18,Syassen08,Tomita17,Diehl08,Poletti12,Poletti13,Diehl08,Sieberer13,Marino16,Schiro16,Minganti18,Rota19,Young20,paz2021drivendissipative,Marcuzzi14,seetharam2021correlation}. The boundary between these phases hosts a rich spectrum of dissipative phase transitions (DPT), that capture how the properties of the `average' state change as a function of a given external parameter. The phases emerging at the onset of a DPT are part of a larger class of out of equilibrium phenomena which can be accessed by suitable non-equilibrium drives, such as those recently reported in various solid state systems \cite{Morrison14,Fausti11,Mitrano16,Zong19,Kogar20,Nova19,Disa20,Chiriaco18,Sentef17,Sun20}.

In parallel to these developments, a series of recent works has introduced a new perspective that, instead of focusing on the properties of the average steady state, studies the many-body properties at the level of single quantum trajectories. In this context, it has been shown how the competition between quantum measurements and coherent dynamics (either analog or digitally generated) can give rise to transitions that manifest themselves in specific observables that are not properties of the averaged state - such as von Neumann entropies, negativities, or two-time correlation functions. This second class of transitions is often referred to as ``measurement induced'' phase transition (MIPT) \cite{Dhar16,Nahum2017, Li18, Li2019,Li20,Li21, Chan19, Zhou19, Skinner2019, Bao2020, Jian20,Gullans2020, Gullans2020a,Gopalakrishnan20,Turkeshi20,Turkeshi2021a,Ippoliti21,Buchhold2021,Minato21,Block2021,Sharma21,Hashizume21, Sahu21, Chen20,Biella2020,Tang21,Zhang21a,Jian21,zhang2021syk, Muller2021,Lunt21,Alberton21,Zhang20,Vasseur19,Lopez-Piqueres20,Nahum21,Cao19,Maity20,Fuji20,Szyniszewski20,Lang15,Vijay20,Iaconis20,Bentsen21,Botzung2021}. While both types of transitions are clear footprints of the competition between coherent and incoherent dynamics, the interplay between them has so far received little attention, in part due to the fact that the two are typically formulated within rather different frameworks. 

\begin{figure*}[!t]
\includegraphics[width=\textwidth]{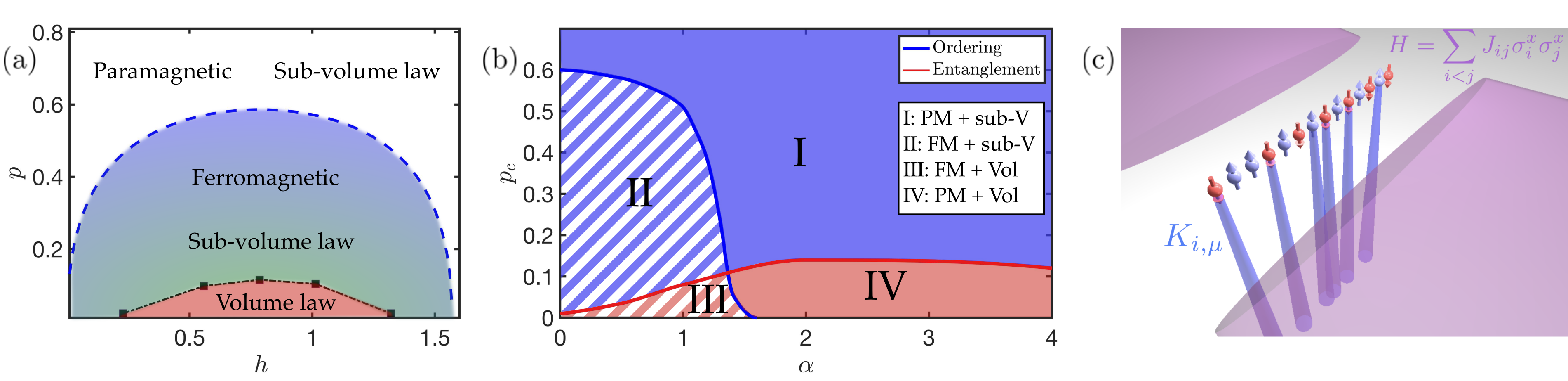}
	\caption{
	Schematic phase diagram and experimental realization. 
	Panel (a) shows the behavior of ordering transition (blue dashed curves) and entanglement transition (black dashed dotted curve) for long-range interactions as a function of resetting probability $p$ and external field $h$. A region of coexistence of the ordered phase and the sub-volume law phase appears for intermediate values of $p$. The dashed blue curve shows the phase boundary calculated according to single site mean field theory. 
	Panel (b) presents a qualitative diagram of the interplay between ordering and entanglement transition as a function of the range of interactions $\alpha$ and the resetting probability $p$. Four types of phases appear: phase I is characterized by paramagnetic disorder (PM) and a sub-volume law scaling of the entanglement; phase II exhibits ferromagnetic order (FM) and sub-volume scaling; phase III shows ferromagnetism and volume scaling, while phase IV is characterized by disorder and volume law. We note that for long-range interactions present in the considered model, our analysis is able to exclude the presence of a volume law, but cannot access system sizes large enough to distinguish between an area law and a sub-volume law; hence we use the latter term to describe the quantum Zeno phase in our model.	Panel (c) outlines a possible experimental configuration for the observation of DPTs and MIPTs with trapped ions. Global off-resonant beams (purple) induce the unitary evolution governed by long-range interactions across the one-dimensional ion chain (as described in Eq. \eqref{Eq:IsHam}), while individual resonant beams (blue) implement the non-unitary operations $K_{i,\mu_i}$ on a randomly selected subset of qubits (see Eq. \eqref{eq:defkrausop}). 
  }
	\label{fig:Sketch_diagram}
\end{figure*}

In this work, we show that many-body dissipative (Liouville) Floquet dynamics provides a natural, generic framework to simultaneously investigate these types of non-equilibrium phase transitions within the same setting. This framework, that has remained largely unexplored so far, enables us to identify both common and distinct features for DPT and MIPT. In particular, while both are typically equally driven by measurement (as both entanglement and quantum order are generally suppressed by off-axis measurements), they do have very distinctive sensitivity to changes in the coherent dynamics: the unitary evolution can in some cases drive an entanglement transition with no  ordered phase, as well as the entire opposite scenario, where a DPT is present while there is no entanglement transition (see Fig.~\ref{fig:Sketch_diagram}a-b).

We illustrate this new framework and our findings utilizing as a concrete model a one dimensional spin system subject to a long-range interacting Hamiltonian evolution interspersed with random quantum measurements~\footnote{Note that dissipative Floquet dynamics has already been used to discuss MITP in Ref.~\cite{Li2019}; however, in those models, no DPT is expected nor has been observed}. The periodic dynamics consists of cycles with a unitary evolution followed by quantum measurements, providing the competition needed to observe the transitions. The unitary evolution is governed by a long-range many-body Hamiltonian with an interaction that decays as a power-law with the distance between particles, and by a transverse external field; the measurement layer consists of random generalized measurement operations which independently reset the spins to the down state with a certain probability.

This setup is very advantageous for the following reasons. Firstly, it allows us to tune both the range of interactions and the strength of measurements at the same time, enabling us to study an interplay of different phenomena which has not been adequately explored in the literature so far. Secondly, this type of transition can be experimentally investigated in a realistic trapped-ion setting; so far experimental realizations of MIPTs have been elusive because a high degree of control on the quantum system is required for both unitary and non-unitary operations on individual qubits. The model we consider is readily realizable in large system sizes \cite{Zhang2017observationDPT} using trapped ions (see Fig.~\ref{fig:Sketch_diagram}c), where the unitary evolution of the long-range 1D spin model can be simulated \cite{monroe2021programmable, tan2021domain,Kyprianidis2021observation}, and local resetting and measurement operations on the spins can be naturally implemented with optical pumping ~\cite{happer1972optical} and state dependent fluorescence~\cite{noek2013high,Christensen2020high}, respectively. 

Within our model, we investigate: (i) an order-disorder DPT, observed when the average state of the system develops a spontaneous magnetization or spin correlations; (ii) a MIPT, where the scaling of the entanglement entropy for the conditional state exhibits a change between volume and sub-volume law, observed also in free theories, random circuits \cite{Nahum2017,Li18,Li2019,Li20,Skinner2019,Turkeshi20,Turkeshi2021a,Ippoliti21,Buchhold2021} and, more recently, in long-range systems \cite{Minato21,Block2021,Hashizume21, Sahu21, Sharma21}.

We show that these transitions do not generally coincide, and that depending on the range of the interactions one type of transition may appear while another one is suppressed. More specifically we find that for long-range interactions the ordering transition differs from the entanglement transition. As the range of the interactions decreases, the ordered phase disappears and only the entanglement transition survives.

Our findings help shed light on the connection between DPT and MIPT, which have been mostly studied in disjoint settings in the literature so far (we note, however, that changes of order across MIPTs has been discussed in Refs.~\cite{Sang20,Bao2021}). Moreover, our results emphasize the usefulness of dissipative Floquet theory, a so-far little employed theoretical framework even in the context of DPT, which are typically framed within the context of time-independent Liouvillians. 

In fact, from the methodological side, we observe that such models lend themselves to accurate analytical computations within (cluster-)mean-field approximations, often recovering even quantitative agreement with numerical simulations \cite{Jin16,Jin18}. The various phases arising from a dissipative Liouvillian dynamics have been studied in several many-body systems, including spin systems \cite{Lee13,Jin16,Jin18,Fink17,Young20} and systems of bosons \cite{Poletti12,Poletti13,Sieberer13}. In particular, spin systems \cite{Lee13,Jin16,Jin18} can display the emergence of an ordered ferromagnetic phase depending on the strength of dissipation, in which correlations between spins may play an important role in determining the phase diagram  \cite{Jin16}.

This paper is organized as follows, in section \ref{Sec:model}, we present a detailed description of the model under consideration. The analytical and numerical methods employed to probe the transitions are detailed in section \ref{Sec:methods}.  In section \ref{Sec:order}, we discuss the order-disorder DPT, followed by an analysis on the purity of the average state; we provide a comparison between the results obtained using mean field and exact diagonalization. The results for the entanglement MIPT are presented in section \ref{Sec:entanglement}. In section \ref{Sec:mulfrac_ss}, the behavior of the participation entropies is discussed as a supporting probe to the different transitions under consideration. Section \ref{Sec:Exp} provides an experimental overview regarding the observation of DPTs and MIPTs, while section \ref{Sec:conc} presents the concluding remarks of the manuscript.

\section{The model}\label{Sec:model}

We consider a one dimensional system of $L$ spin-$1/2$ interacting between them via a long-range Ising interaction and subject to a transverse external field. The unitary evolution and the measurement processes are periodic in time with period $T$.

\begin{figure}
	\includegraphics[width=1\columnwidth]{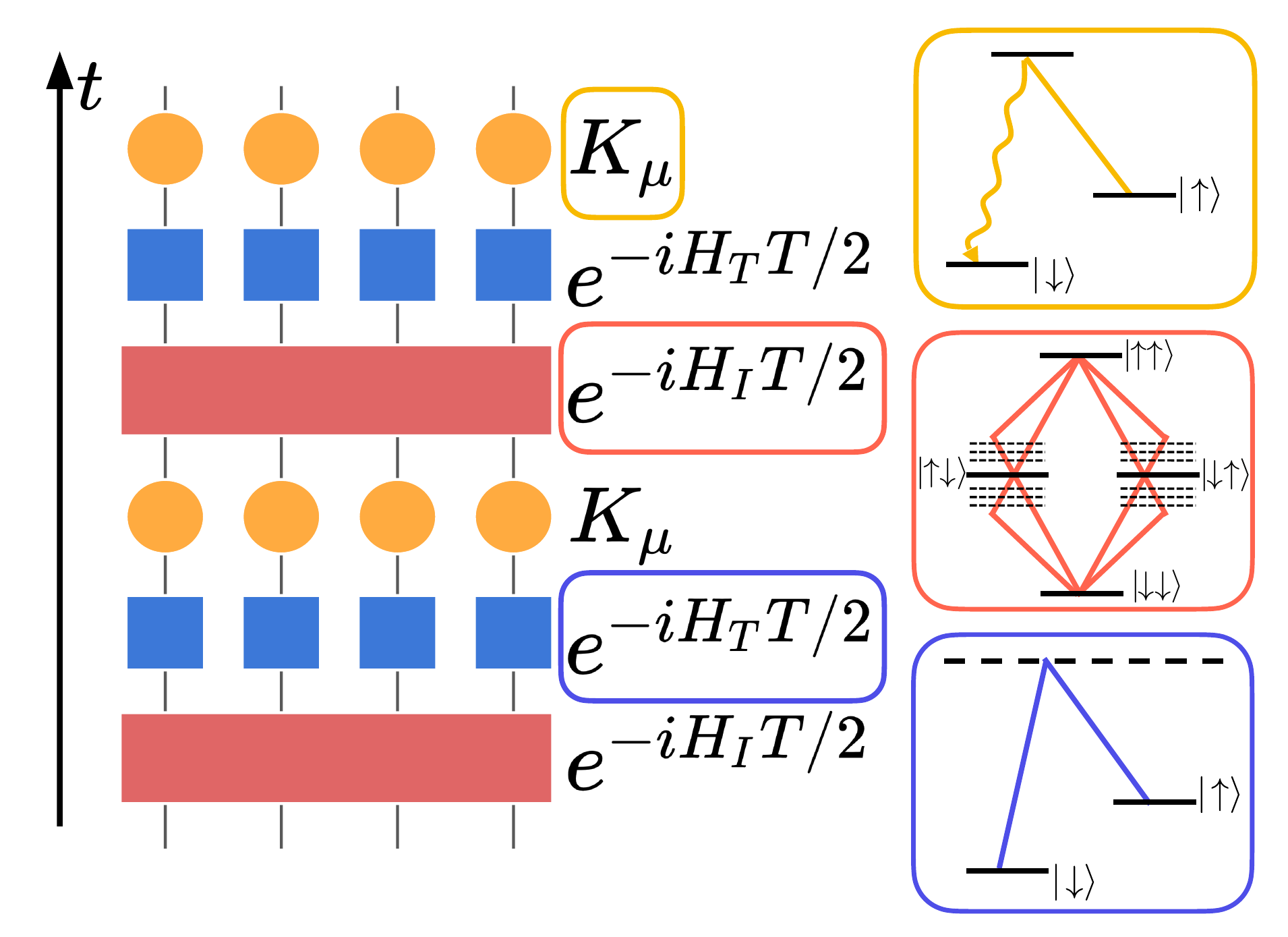}
	\caption{{Scheme of the evolution protocol of the system. The red layer represents the evolution step with the long-range interaction Hamiltonian $H_I$, which couples all the spins mediated by motional modes of the ion crystal \cite{Molmer1999,monroe2021programmable} (dashed black lines). The blue layer represents the evolution with the transverse field Hamiltonian $H_T$ coherently coupling the two spin states with a two photon process (see section \ref{sec:ExpA}). Finally, the yellow circles layer corresponds to the measurement process, implemented with resonant photon scattering.}
  }
	\label{fig:tevol}
\end{figure}

The dynamics of systems subject to measurements continuous in time can be effectively described by a quantum master equation for the average density matrix $\rho$ \cite{Minganti18,Gopalakrishnan20,Lee13,Schnell20}. However we consider discrete measurements occurring on a very short timescale at times $t=nT$, so that the unitary dynamics and the measurements occur sequentially rather than simultaneously. Within this setting, a description of the dissipative dynamics through Lindblad operators is not very practical; instead we will be using a Kraus map to describe the measurement process (see Fig.~\ref{fig:tevol}).

\subsection{Hamiltonian evolution}\label{Sec:modelHam}

We first define the Hamiltonian governing the unitary dynamics. For practical reasons that will be explained in detail in Section \ref{Sec:methods}, we assume that the unitary evolution occurs in a ``kicked'' fashion: the system is subject only to the Hamiltonian $H_I$ of the long-range interaction for times $0<t-nT<T/2$ and then only to the external field Hamiltonian $H_T$ for times $T/2<t-nT<T$.

Similarly to any Floquet dynamics, we can write the evolution operator $\mathcal U$ over one cycle \cite{Bukov15}
\begin{gather}\notag
\mathcal U=e^{-iH_T T/2}e^{-iH_IT/2};\\
\label{Eq:IsHam}
H_I\equiv-\sum_{i\neq j}J_{ij}\sigma_i^x \sigma_j^x;\qquad
H_T\equiv h\sum_i \sigma_i^z,
\end{gather}
where $\sigma_i^a$ are the Pauli matrices on site $i$ and $h$ is the external field. The long range interaction in $H_I$ assumes periodic boundary conditions and includes a Kac normalization
\begin{equation}\label{Eq:KacN}
\begin{split}
J_{ij}=\frac{J}{\mathcal{N}} \left( \frac1{|i-j|^{\alpha}} + \frac1{(L-|i-j|)^{\alpha}} \right), \\
\mathcal{N}\equiv\sum_{r=1 }^{L-1}\left( \frac1{r^{\alpha}}+\frac1{(L-r)^{\alpha}} \right).
\end{split}
\end{equation}

The quantity $\alpha$ parameterizes the range of the interaction. In equilibrium, interactions with $0<\alpha<1$ are typically considered to be long range, while values $\alpha\gtrsim2$ yield short range interactions \cite{Zunkovic16}.
The long-range interactions lead to a sub-linear growth of entanglement entropy after a quench \cite{Hauke13, Schachenmayer13,lerose1,silvia1,lerose2}.

For all numerical calculations from now on we set $J=1$ and $T=1$.

\subsection{Quantum measurements}\label{Sec:modelDis}

As already mentioned, the measurement process can be described by a set of Kraus operators $\{K_{\mu}\}$ which can be related to the Lindblad jump operators. The Kraus operators preserve the global $\mathbb{Z}_2$ symmetry of the Ising model, i.e. $K_\mu (U_Z \rho U_Z^\dagger) K_\mu^\dagger=U_Z (K_\mu \rho  K_\mu^\dagger) U_Z^\dagger$ with $U_Z=\prod_i \sigma_i^z$, 
for every $\mu$ and every $\rho$, and satisfy the normalization condition $\sum_\mu K_\mu^\dagger K_\mu=1$.

We consider a resetting measurement process, where spins are independently reset to the down state with probability $p$. 
The Kraus operators associated with these generalized measurements $\mathbb{E}_\mu \equiv K^\dagger_\mu K_\mu$ are labeled by a multi-index $\mu=(\mu_1,\dots, \mu_L)$, with $\mu_i=0,1,2$ and are defined as \footnote{Notice that this is one choice of Kraus operators for the resetting process; other choices are possible and have the same action on the average state of the system.}
\begin{equation}
\label{eq:defkrausop}
\begin{split}
    K_\mu&=\prod_i K_{i,\mu_i}\\ 
    K_{i,0} =\sqrt{p} \ket{\downarrow}\bra{\downarrow}_i, &\qquad K_{i,1} =\sqrt{p} \ket{\downarrow}\bra{\uparrow}_i, \\
    K_{i,2}&= \sqrt{1-p}\; \mathbf{1}_i.
\end{split}
\end{equation}

Depending on the observable we want to investigate, we can limit ourselves to the study of the average state, given by the density matrix $\rho=\overline{\ket{\psi}\bra{\psi}}$, or we may have to consider single quantum trajectories
\cite{Plenio98, Carmichael09, Daley14}. 

At the level of a quantum trajectory, each measurement on a single spin corresponds to a sudden change in the state of the system $\ket{\psi}\rightarrow\ket{\psi'}$ with
\begin{equation}
    \label{Eq:trajectory_dis}
    |\psi'\rangle = \frac{K_{\mu} |\psi\rangle}{\sqrt{\langle \psi|K^\dagger_{\mu} K_{\mu}|\psi\rangle }}.
\end{equation}
The choice of which Kraus operator is applied for each site is given by the Born rule via the probability distribution $\mathcal{P}(\mu)=\langle \psi|\mathbb E_{\mu}|\psi \rangle=\prod_i \langle \psi|K^\dagger_{i,\mu_i} K_{i,\mu_i}|\psi\rangle$.

The action of the dissipative map on the average state can be found by summing over the outcomes of the measurement
\begin{gather}\label{Eq:Kraus}
\rho' = \sum_\mu K_\mu \rho K_\mu^\dagger;\\
\label{Eq:rho_meas}
    \rho'= p \ket{\downarrow}\bra{\downarrow}_i\otimes\text{Tr}_i\rho+ (1-p) \rho.
\end{gather}
where $\rho$ ($\rho'$) is the density matrix of the system before (after) the measurement, and $\text{Tr}_i$ denotes the partial trace over the degrees of freedom associated to the spin $i$.

Combining Eqs.~\eqref{Eq:Kraus}-\eqref{Eq:rho_meas} with the unitary evolution \eqref{Eq:IsHam} it is possible to write a quantum master equation for the average state
\begin{equation}\label{Eq:MasterEq}
\partial_t\rho=\mathcal{L}(t)[\rho],
\end{equation}
where the Lindblad superoperator $\mathcal{L}(t)$ is a function of the Hamiltonian and of the Kraus operators and is periodic in time. Master equations of the form \eqref{Eq:MasterEq} are routinely used to investigate DPTs, but are not very suited to analyze a single trajectory.

Crucially, such analysis of the quantum trajectories is needed when studying observables that are non-linear in the state of the system.
Examples relevant for this paper are the entanglement entropy or the purity. As we shall see, for these quantities the conditional average over the trajectories is different from the analog one computed on 
the average state \footnote{Here by conditional average we mean computing first the expectation value of the observable over the state and then taking the average over the trajectories.}, and for this reason they provide a sensible quantity for the MIPT.

After a certain number of cycles (depending on the value of $p$) of unitary evolution plus measurement, the system reaches a steady state, in the sense that all the properties of the state (expectation values of observables, entanglement entropy, etc.) averaged over trajectories reach stationary values. We note that these quantities are not necessarily stationary during one cycle since the unitary dynamics and the measurement may still change them, but they are constant when considering two equivalent times in different cycles. For example, the steady state average density matrix is $\rho_\mathrm{ss}(t)=\lim_{n\rightarrow\infty}\overline{\ket{\psi(t+nT)}\bra{\psi(t+nT)}}$ and satisfies $\rho_\mathrm{ss}(t+T)=\rho_\mathrm{ss}(t)$.

\section{Methods}\label{Sec:methods}

We aim to characterize the properties of the steady state by studying the magnetization (and the spin correlators),  entanglement properties, and purity. To that end, we combine analytical methods with numerical simulations.

In this section we give an overview of the methods we employ to investigate the relevant observables, which are reported in Table \ref{tab:my_label}.

\subsection{Mean field approximation}\label{Sec:methodsMF}

The analytical methods we utilize are based on single site mean field and cluster mean field approximations. While those methods cannot be informative about bipartite entanglement properties, they have access to both local order parameters and the purity of the system. Moreover, their simplicity offers several simple insights on the dynamics of the system in a regime where the approximation is very well controlled, as discussed below. This is expected, given the impressive predictive power of similar methods in the context of quantum quench dynamics for spin models with $\alpha<1$ \cite{Das2006,Sciolla2010,snoek2011rigorous,sciolla2011dynamical}.

\emph{Single site mean field approximation} - We can study the evolution of the system in a single site mean field approximation. Within such approximation, each spin is decoupled from the others, i.e. the density matrix factorizes $\rho=\bigotimes_i\rho_i$.

We can write down the evolution equations that determine the unitary dynamics of the average value of the spin vector $\vec s_i\equiv\overline{\ave{\vec{\sigma}_i}}=\text{Tr}(\vec{\sigma}_i\rho_i)$; $\overline{\cdot}$ indicates the average over trajectories, while $\ave{\cdot}$ is the quantum expectation value. Note that the trace over $\rho$ automatically takes into account the average over all quantum trajectories. The equations of motion are
\begin{gather}\label{Eq:MFdyn}
\frac{d}{dt}\vec{s}_i=-\vec B_{MF}\times\vec{s}_i;\\
\label{Eq:MFspin_B}\vec B_{MF}\equiv \begin{cases}
               4\sum_{k\neq i}J_{ik}s^x_k\hat x, \quad 0<t<T/2\\
               -2h\hat z, \quad T/2<t<T
            \end{cases}
\end{gather}
where $\hat{i}$ is the versor in the direction $i=x,y,z$.
Equations \eqref{Eq:MFdyn}-\eqref{Eq:MFspin_B} are to be solved self consistently since the effective field $\vec B_{MF}$ depends non linearly on $\vec s_i$ during the first half of the unitary evolution cycle.

From Eq. \eqref{Eq:rho_meas} we find the action of the dissipative map on the spins:
\begin{equation}\label{Eq:MFKraus}
s^{x,y}_i\rightarrow(1-p)s^{x,y}_i; \quad s^z_i\rightarrow(1-p)s^z_i-p.
\end{equation}

\emph{Cluster mean field approximation} - The single site mean field approximation completely neglects any correlation between spins. A simple way to include some degrees of correlations is to consider a cluster mean field approximation, in which the density matrix factorizes as the product of the density matrices of neighboring spins $\rho=\bigotimes_i\rho_{2i-1,2i}$. We consider the density matrix of the first two sites $\rho_{12}$.

The mean field Hamiltonian reads
\begin{gather}
\label{Eq:ClusterMFHI}
H_I^{MF}=-2J\left(1-\frac1{\mathcal{N}}\right)s^x(\sigma_1^x+\sigma_2^x)-\frac {2J}{\mathcal{N}}\sigma_1^x\sigma_2^x;\\
\label{Eq:ClusterMFHT}
H_T^{MF}=h(\sigma_1^z+\sigma_2^z);\\
\label{Eq:ClusterMFsx}
s^x=\frac12\langle\sigma_{1}^x+\sigma_2^x\rangle=\frac12\text{Tr}[\rho_{12}(\sigma_{1}^x+\sigma_2^x)].
\end{gather}

The action of the dissipative map on the density matrix can be obtained from Eq. \eqref{Eq:Kraus}
\begin{align}
\notag\rho_{12}\rightarrow&(1-p)^2\rho_{12}+p(1-p)\ket{\downarrow}\bra{\downarrow}_1\text{Tr}_1\rho_{12}+\\
\label{Eq:clusterMFKraus}
+&p(1-p)\ket{\downarrow}\bra{\downarrow}_2\text{Tr}_2\rho_{12}+p^2\ket{\downarrow}\bra{\downarrow}_1\ket{\downarrow}\bra{\downarrow}_2.
\end{align}

Notice that the terms proportional to $p(1-p)$ involve resetting of only one of the spins, and that evaluating the partial trace over one of the spins creates a mixed state if the density matrix does not correspond to a pure state with both spins in the down state.

The unitary evolution is determined by 
\begin{equation}\label{Eq:clusterMFdtrho}
\frac{d\rho_{12}}{dt}=-i
\begin{cases}
[H_I^{MF},\rho_{12}]\, \text{ for } \, 0<t<T/2,\\
[H_T^{MF},\rho_{12}]\, \text{ for } \, T/2<t<T.
\end{cases}
\end{equation}

Note that for $\alpha\le 1$, the two spins of the cluster --- whose interaction is contained in Eq. (\ref{Eq:ClusterMFHI}) --- are completely decoupled in the thermodynamic limit because $1/\mathcal{N}\rightarrow 0$. Therefore, cluster mean field coincides with single site mean field for $\alpha\le 1$ (and we will generically refer to it as ``mean field"), while for $\alpha>1$ it gives more accurate predictions because it includes nearest-neighbor correlations.

\subsection{Exact numerical simulations}\label{Sec:methods_ED}

We also utilize extensive exact simulations that are best suited to tackle entanglement dynamics and provide quantitative predictions for the transition properties. Within our dissipative Floquet setting, it is of pivotal importance to access as large system sizes as possible with full numerical control. For this purpose, we utilize the following numerical procedure.

To start, we note that $\mathcal U$ is a product of operators that are diagonal respectively in the eigenbases of $ \sigma^x_i$ and $\sigma^z_i$. 
The multiplication of a vector by a diagonal matrix requires only $N=2^L$ operations as opposed to $N^2$ operations for a multiplication by a full matrix. To utilize this fact, we also note that the transformation between the eigenbases of the $\sigma^x_i$ and $\sigma^z_i$ operators is given by $R=\bigotimes_{i=1}^L R_i$, where $R_i=\frac{1}{\sqrt{2}} \big(\begin{smallmatrix}
  1 & 1\\
  1 & -1
\end{smallmatrix}\big)$. The product structure of this transformation allows us to use an approach analogous to fast Fourier transformation, known as the fast Hadamard transform \cite{Fino76, Arndt11} which requires only $N \log N \sim L 2^L$ operations to transform between eigenbases of $ \sigma^x_i$ and $ \sigma^z_i$ operators. Hence, the multiplication of a vector $\ket{\psi}$ by the matrix $\mathcal U$ reduces to two fast Hadamard transforms and two multiplications by the diagonal matrices $ e^{-iH_TT/2}$ and $ e^{-iH_IT/2}$, resulting in a total time complexity $L 2^L$. Indeed, similar approaches have been employed in studies of the ergodicity of many-body systems \cite{Prosen98, Prosen99, Lezama19}. Finally, the measurement \eqref{Eq:trajectory_dis} can be represented by the action of sparse operators in the $\sigma^z_i$ basis. This allows us to simulate exactly the time evolution of systems consisting of up to $L=24$ spins within reasonable computing times.

This procedure shows why it is advantageous to consider a kicked Ising Hamiltonian in comparison to a traditional Ising Hamiltonian that contains both the interaction term and the transverse field term at the same time.

\begin{table}[]
    \centering
    \begin{tabular}{c|c|c}
        Observable & Transition & Definition \\
        \hline
        $X(t)$ & Ordering & Eq. \eqref{eq:X}\\
        $X^2(t)$ & Ordering & Eq. \eqref{eq:XX}\\
        $\mathcal B$ & Ordering & Eq. (\ref{Eq:Binder})\\
        QMI & Entanglement & Eq. \eqref{Eq:qmi}\\
        $S_R$ & Entanglement & Eq. \eqref{Eq:RefQubS}
    \end{tabular}
    \caption{\scriptsize{Table of the observable used to study the different transitions.}}
    \label{tab:my_label}
\end{table}

\subsection{Exact numerics for $\alpha=0$}\label{Sec:methods_SYMM}
The limit $\alpha=0$ for the average evolution Eq.~\eqref{Eq:Kraus} can be effectively simulated due to an emerging permutation symmetry. In fact, in this limit the interactions in the Hamiltonian have infinite range: we can use this property to perform our numerical simulations in permutational symmetric subspaces, allowing to study larger system sizes.

To explain how this is done, it is convenient to work in the superoperator formalism, transforming the density matrix to a vector with the following mapping
\begin{equation}
\label{eq:superoperator}
    \rho=\sum_{i,j} \rho_{i,j} \ket{i}\bra{j}\; \Longrightarrow\;\ket{\rho}\rangle =\sum_{i,j}\rho_{i,j}\ket{i}\ket{j}.
\end{equation}
The vector $\ket{\rho}\rangle$ is hence defined on a ``doubled" system, with $L$ sites associated with the left part (the state $\ket{i}$ in Eq.~(\ref{eq:superoperator})) and $L$ sites for the right part (the state $\ket{j}$).

The time evolution of the vector $\ket{\rho}\rangle$ induced by the unitary dynamics with an Hamiltonian $H(t)$ and by the measurement of Eq.~(\ref{Eq:rho_meas}) is 
\begin{gather}
\label{Eq:superopEvolution}\dot{\ket{\rho}}\rangle= -i\Big(H(t)\otimes 1-1\otimes H(t)^*\Big)\ket{\rho}\rangle;\\
\label{Eq:superopDissipation}\ket{\rho}\rangle'= \sum_\mu K_\mu \otimes K_\mu^* |\rho\rangle\rangle.
\end{gather}

By using the explicit form of the Kraus operators in Eq.~(\ref{Eq:Kraus}) we get
\begin{equation}\label{Eq:superopKraus}
    \sum_\mu K_\mu \otimes K_\mu^*=\prod_i\sum_{\mu_i=0}^2 K_{i,\mu_i} \otimes K_{i,\mu_i}^*
    =\exp\left(\sum_i M_i\right),
\end{equation}
where the operator $M_i$ acts on the left and right spins at position $i$ and has the form
\begin{equation}\label{Eq:superopKrausM}
    M_i= -\ln (1-p)\Big( \ket{\downarrow\downarrow}\bra{\downarrow\downarrow}+\ket{\downarrow\downarrow}\bra{\uparrow\uparrow}-\mathbf{1}\otimes \mathbf{1}\Big)_i.
\end{equation}

Note that, while for $\alpha=0$ the Hamiltonian part is invariant under both the permutations of the $L$ sites on the left and the permutations of the $L$ sites on the right (the symmetry is given by the group $\mathcal{S}_L^{(l)}\times \mathcal{S}_L^{(r)}$), the measurement is only invariant under permutations of the left and right part simultaneously (the symmetry group is $\mathcal{S}_{L}^{(l,r)}$). We define a basis for the invariant subspace under these simultaneous permutations: each state of the basis is labeled by a sequence of four integers $\{n_i\}_{i=1,\dots,4}$ with $0\le n_i \le L$ and $\sum_i n_i=L$, and is defined as
\begin{equation}
\label{eq:perm_basis}
    \ket{\{n_i\}}=\frac{1}{\mathcal{N}_{\{n_i\}}}\sum_{\tau\in \mathcal{S}_L}\mathbb{S}_\tau \ket{\uparrow\uparrow}^{\otimes n_1}\ket{\uparrow\downarrow}^{\otimes n_2}\ket{\downarrow\uparrow}^{\otimes n_3}\ket{\downarrow\downarrow}^{\otimes n_4}
\end{equation}
where $\mathcal{N}_{\{n_i\}}$ is a normalization constant, the sum runs over all the permutations $\tau$ of $L$ elements and $\mathbb{S}_\tau$ is the operator that applies the permutation $\tau$ to the $L$ (doubled) sites of the system. The dimension of this subspace is $\binom{L+3}{3}$: the slow polynomial scaling in $L$, compared with the exponential scaling $2^{2L}$ of the full Hilbert space of the density matrix, allows us to simulate the exact dynamics of systems comprised of few hundreds spins. In Appendix \ref{sec:app_perm} we report the calculation of the matrix elements of the operators associated with $H_T$, $H_I$ and of the operator $M=\sum_i M_i$ in this basis, and the procedure for computing expectation values.
We note that similar approaches have been used for various open quantum systems with collective incoherent processes \cite{Sarkar87,
Chase08, Baragiola10, Lee12, Xu13, Marduk15, Gegg16, Gegg17, Gegg18, Shammah18, Iemini18,Hama18}.

\section{Order-disorder transition}\label{Sec:order}

In this section we demonstrate that the steady state of the kicked Ising model with random measurement hosts a symmetry breaking transition, similar to ordering transitions in ground states of quantum magnets \cite{Sachdev11}.
The ordering transition is best revealed by order parameters such as the spontaneous magnetization of the spins in the $x$ direction or suitable correlation functions. We study the average magnetization $X$ (averaged over trajectories and spins) and 
the two point correlation function $X^2(t)$:
\begin{gather}
\label{eq:X}
X(t)=\frac1L\sum_{i=1}^L\overline{\ave{\psi(t)|\sigma^x_i | \psi(t)}};\\
\label{eq:XX}
X^2(t)=\frac1{L(L-1)}\sum_{i\neq j}^L\overline{\ave{\psi(t)| \sigma^x_i\sigma^x_j | \psi(t)}},
 \end{gather}
 where $\overline{A}$ is the average of the quantity $A$ over the many quantum trajectories that arise when the various measurements \eqref{Eq:trajectory_dis} at random positions and times yield different outcomes.

The ordered phase of our model is characterized by a non-vanishing   
 magnetization and   two-point correlation function, whereas in the disordered phase 
 $X=0=X^2$. 
The two phases arise from the competition between the unitary time evolution \eqref{Eq:IsHam} and the resetting process. The unitary time evolution enhances the ferromagnetic correlations in the steady state for $J>0$ and sufficiently small $h>0$. In contrast, the resetting process favors a paramagnetic state \cite{Rossini20}. Indeed, 
in the limit $p\rightarrow1$, all spins are always reset, so that the steady state is trivially $|\!\!\downarrow\,\downarrow\,...\,\downarrow\rangle$. Notice that all correlations in this state are destroyed by the measurement process, and that the value of each spin is solely determined by the measurement on that spin \footnote{For example if the first spin were subject to a resetting to the up state, the system steady state would be $|\uparrow\,\downarrow\,...\,\downarrow\rangle$.}.

We observe that the average over trajectories and the expectation value in Eqs. 
\eqref{eq:X}-\eqref{eq:XX} commute because we are dealing with quantities that are linear 
in the average steady state of the system $\rho_{ss}(t)$. Indeed, the magnetization averaged over quantum trajectories \eqref{eq:X} and the magnetization of the average state coincide $X= \frac{1}{L}\text{Tr}( \rho_{ss}  \sum_{i}\sigma^x_i)$,  and similarly for $X^2(t)$.
Therefore, an analysis based on the evolution of the density matrix and an analysis studying the single trajectories yields the same results.

To study the symmetry breaking transition we first investigate the kicked Ising model with resetting in the mean field approximation and then demonstrate that the findings of mean field analysis are consistent with the properties of the steady state derived from numerical simulations. We also highlight the crucial role played by the range of interaction in stabilizing the long-range order in the considered system.

\subsection{Mean field analysis}\label{Sec:orderMF}

We start our investigation of the symmetry breaking transition using the single site mean field approximation  of Section \ref{Sec:methodsMF} -- which is expected to work well in the infinite range $\alpha\rightarrow0$ regime~\cite{sciolla2011dynamical}. 

From Eq. \eqref{Eq:MFKraus} we note that the spin component in the $x-y$ plane is exponentially suppressed by the measurements, while it may increase during the unitary evolution. This means that when the probability of measurement is very high, the steady state is a disordered phase characterized by the fixed point $\vec{s}_i=-\hat z$ of the evolution map. On the other hand for smaller $p$ the system reaches an ordered phase characterized by the non zero ``order parameter'' $s_i^{x}\neq 0$.

Equations \eqref{Eq:MFdyn}-\eqref{Eq:MFKraus} can be solved numerically in general. However, we can perform an analytical treatment close to the boundary between ordered and disordered phase, where the order parameter is very small, i.e. $0<|s_i^{x,y}|\ll1$. We also assume that there is no site dependence of the spin and write $s_i^{x}(t)=\epsilon^x(t)$, $s_i^{y}(t)=\epsilon^y(t)$, $s_i^{z}(t)=-1+\epsilon^z(t)$; we can then linearize Eq. \eqref{Eq:MFdyn} and find
\begin{gather}\label{Eq:MFlinX}
\dot\epsilon^x=\epsilon^y\begin{cases}
0, \quad 0<t<T/2\\
-2h, \quad T/2<t<T
\end{cases};\\
\label{Eq:MFlinYZ}\dot\epsilon^y=\epsilon^x\begin{cases}
-4J, \quad 0<t<T/2\\
2h, \quad T/2<t<T
\end{cases};\quad
\dot\epsilon^z = 0; 
\end{gather}

Equations \eqref{Eq:MFlinX}-\eqref{Eq:MFlinYZ} can easily be solved given the initial conditions $(\epsilon^x,\epsilon^y,\epsilon^z)=(A,B,C)$. The fixed point conditions $(1-p)\epsilon^{x,y}(t=T^-)=\epsilon^{x,y}(t=0^+)$, $(1-p)(-1+\epsilon^{z}(t=T^-))-p=(-1+\epsilon^{z}(t=0^+))$ then read
%
\begin{equation*}
\overline{p}
\begin{pmatrix}
\cos(hT)+2JT\sin(hT) & -\sin(hT)\\
\sin(hT)-2JT\cos(hT) & \cos(hT)
\end{pmatrix}
\begin{pmatrix}
A\\
B
\end{pmatrix}=
\begin{pmatrix}
A\\
B
\end{pmatrix},
\end{equation*}
where $\overline p \equiv 1-p$ and $C=0$.

A non-trivial solution exists when
\begin{equation}\label{Eq:MFlinpc}
(1-p)^2-2(1-p)\left[\cos (hT)+JT\sin (hT)\right]+1=0.
\end{equation}

The solution of Eq. \eqref{Eq:MFlinpc} gives the critical probability $p_c(h,J)$ as function of the transverse field $h$ and the interaction strength $J$: for $p>p_c$ there is a single stable fixed point $\vec s_i=-\hat z$ and the system is in the disordered phase; at $p=p_c$ there is bifurcation and the fixed point $\vec s_i=-\hat z$ becomes unstable, while two new stable fixed points appear; the new stable fixed points are characterized by $\overline{\ave{\sigma_i^x}}\neq0$, hence for $p<p_c$ the steady state is ordered.

\begin{figure}
\vspace{0.25cm}
	\includegraphics[width=1\columnwidth]{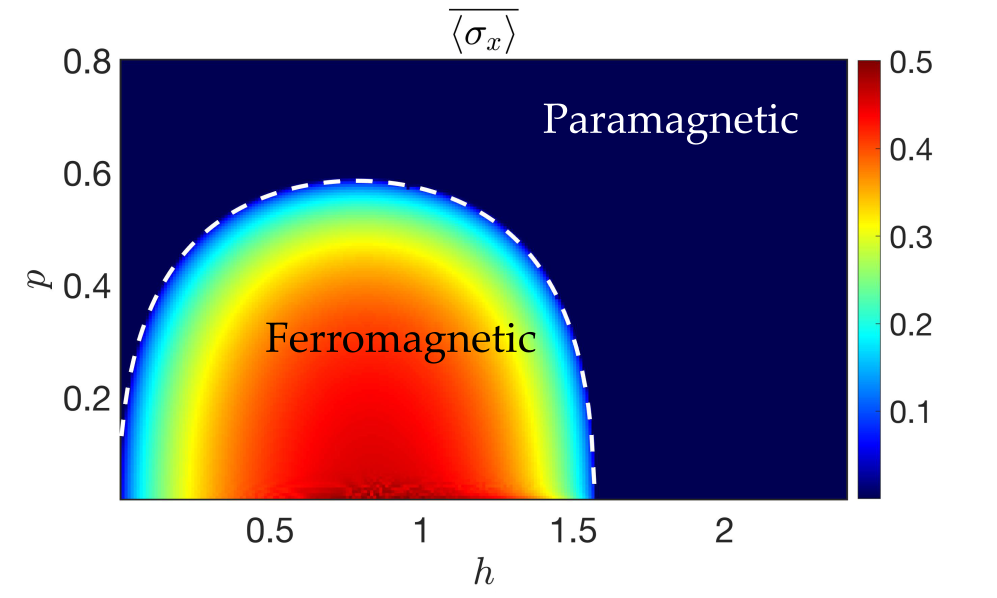}
	\caption{ {Order-disorder transition in the mean field approach. The color map shows the behavior of the absolute value of the average magnetization $\overline{\langle\sigma^x\rangle}$ in the $x$ direction computed in a mean field approximation as function of $h$ and $p$. The white dashed line reports the contour for which $|\overline{\langle\sigma^x\rangle}|=0.01$ which matches the solution of Eq.~\eqref{Eq:MFlinpc}.
	}}
	\label{fig:Color_map1}
\end{figure}

The phase boundary given by $p_c$ for $J=1$, $T=1$ has a lobe shape in the $h-p$ space and matches the results obtained by solving numerically the mean field equations \eqref{Eq:MFlinX}-\eqref{Eq:MFlinYZ}, as reported in Fig. \ref{fig:Color_map1}.

\subsection{Numerical simulations of the steady state}\label{Sec:orderED}

To verify the conclusions of the mean field analysis, we perform exact numerical simulations using the techniques outlined in \ref{Sec:methods_ED} and \ref{Sec:methods_SYMM}.

We take a ferromagnetic state $\ket{\psi_0}=\ket{ \rightarrow \rightarrow ...  \rightarrow  }$ polarized in the $x$ direction as the initial state (we checked explicitly that the results reported here do not depend on the choice of the initial state), and calculate the time evolved state $\ket{\psi(t)}$ by repeated applications of the unitary operator $\mathcal{U}$ in Eq. \eqref{Eq:IsHam} followed by the resetting operation specified in Eq. \eqref{Eq:trajectory_dis}. We then calculate the values of the correlation functions $X(t)$ and $X^2(t)$.

\begin{figure}[!t]
	\includegraphics[width=1\columnwidth]{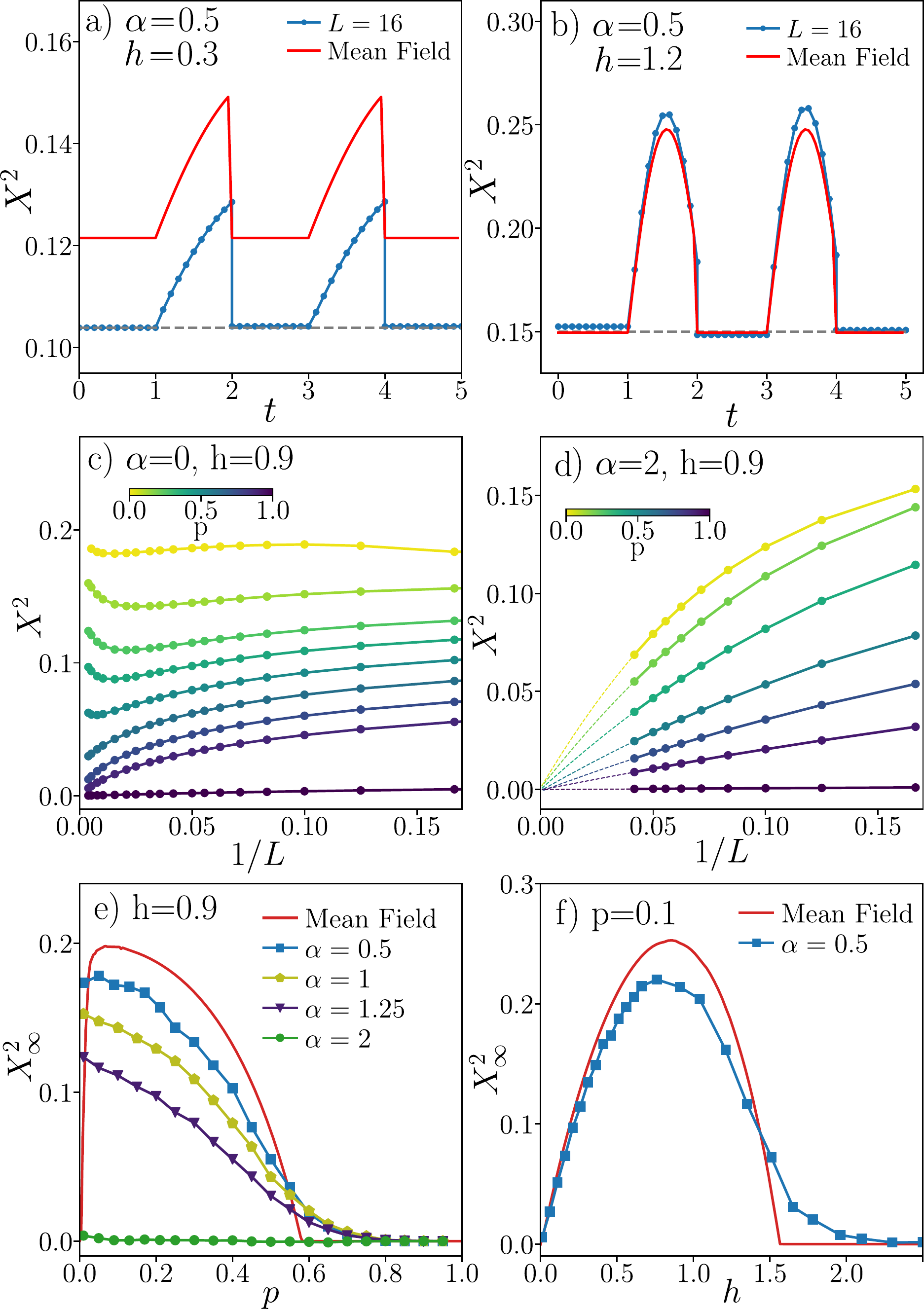}
	\caption{The two-point correlation function $X^2$ in the kicked Ising model with resetting. Panels a), b) show the time dependence of the two-point correlation function $X^2(t)$ in the steady state of the system. The steady state value of $X^2$ at $t=nT+0^+$ is indicated by the dashed lines. Panels c), d) show the average value of the correlation function $X^2$ (measured just after the resetting) in the steady state for various resetting probabilities $p$ and system sizes $L$ respectively for $\alpha=0$ and  $\alpha=2$. Panels e) and f) show the value of correlation function $X^2_{\infty}$ extrapolated to thermodynamic limit $L\rightarrow  \infty$ compared with the mean field (red line) 
	as functions of $p$ and $h$, respectively. 
  }
	\label{fig:phase}
\end{figure}

We observe that the system reaches a steady state for times $t >  t_{\mathrm{steady} }=\max\{ 2L, 10/p\}$, after which the two-point correlation function $X^2(t)$ does not change between subsequent intervals of time $T$, i.e. $X^2(t+T)=X^2(t)$. Examples of time dependence of $X^2(t)$ within a few cycles of evolution of the system are shown in Fig.~\ref{fig:phase}~a),~b). During the first stage of the unitary evolution, $t\in[nT,nT+T/2]$ (where $n$ is an integer), the system evolves with the Hamiltonian $H_I$ which does not change the value of $X^2(t)$. The value of $X^2(t)$ changes in the interval $t\in[nT+T/2, nT+T]$ when the system evolves with the Hamiltonian $H_T$. Finally, upon the resetting process at time $t=nT+T$, the value of $X^2(t)$ jumps discontinuously. This behavior is visible both for the value of $X^2(t)$ calculated for an infinite system size in the mean field approximation as well as from the value yielded by the exact numerical simulations of a system of finite size $L$.

The behavior of the magnetization $X(t)$ is slightly different. While $X(t)$ acquires a steady-state value in mean field approximation, it decays approximately exponentially in time $X(t) \propto e^{-\gamma(p, L) t}$ in a finite system of size $L$. The constant of decay $\gamma(p,L)$ decreases with increasing system size $L$ with rate that depends on the resetting probability $p$. This means that both in the exact numerical simulations of system of size $L$, as well as in a real experimental situation, the steady state value of the two point-correlation function $X^2$ is a more convenient probe of the symmetry breaking transition in our model.

To probe the symmetry breaking transition, we consider the value of the two-point correlation function $X^2(t_n)$ at times $t_n=nT+0^+$ (i.e. just after the resetting), and perform the average over $n$ for $1000T>t_n> t_{\mathrm{steady} }$ to smooth out the small cycle to cycle residual fluctuation of $X^2(t_n)$. For each system size $L$ we consider, we find a non-zero value of the two-point correlation function $X^2$. For the system with interactions of infinite range, $\alpha=0$, we utilize the $\mathcal{S}_L^{(l,r)}$ symmetry to calculate the average steady state $\rho_{ss}(t)$ by exact numerical simulation of Eqs. \eqref{Eq:superopEvolution} and \eqref{Eq:superopDissipation} in the symmetric subspace. This allows us to investigate systems of size of up to $L=384$ sites. The resulting values of $X^2$, shown in Fig.~\ref{fig:phase}c), reveal that the average two-point correlation function is a non-monotonous function of the system size $L$: when $p<p^{SB}_c$, where $p^{SB}_c$ is the critical resetting probability, $X^2$ increases with system size beyond a certain system size $L_0$. The length scale $L_0$ is enlarged when the resetting probability $p$ gets closer to the critical value $p\rightarrow p^{SB-}_c$. In contrast, for $p>p^{SB}_c$, the two-point correlation function $X^2$ decreases monotonously with $L$, which suggests that it vanishes in the thermodynamic limit.

The results for $\alpha=0$ indicate the presence of a symmetry breaking transition in the kicked Ising model. When $\alpha>0$, the average steady state $\rho_{ss}(t)$ is no longer $\mathcal{S}_L^{(l,r)}$ invariant, and to probe the symmetry breaking transition we calculate the average two-point correlator performing the average \eqref{eq:XX} over quantum trajectories which allows us to investigate systems consisting of up to $L=24$ spins. Results for $\alpha=2$ are shown in Fig.~\ref{fig:phase}~d). We immediately note that the decay of $X^2$ with the system size is much more abrupt for the system with interactions of shorter range ($\alpha=2$) as compared to the $\alpha=0$ case. To investigate this quantitatively, we fit $X^2$ with a second order polynomial
\begin{equation}\label{Eq:X2ansatz}
X^2=X^2_{\infty}+a/L+b/L^2.
\end{equation}
The resulting values of $X^2_{\infty}$, shown in Fig.\ref{fig:phase}~e),~f) are the main results of this section. They indicate a transition between an ordered phase in which $X^2_{\infty}>0$, and disordered phase with vanishing value of the extrapolated two-point correlation function. This transition is observed both as a function of the resetting probability $p$ and of the field $h$.

The extrapolated values $X^2_{\infty}$ of two-point correlation functions can be compared with the predictions of mean field approximations, denoted by  red lines in Fig.\ref{fig:phase}~e),~f). Despite the crude form of the ansatz \eqref{Eq:X2ansatz}, the agreement of $X^2_{\infty}$ for $\alpha=0.5$ with the value of the correlation function obtained in the mean field approximations is very good. The discrepancies at the values of $p$ and $h$ around the transition (at which the order parameter vanishes) are in part due to the non-zero value of $\alpha$ (the mean field approaches are expected to be most accurate for $\alpha\rightarrow 0$) and in part to finite size effects which are more severe close to the transition. The presence of the finite size effects that cannot be captured within the ansatz \eqref{Eq:X2ansatz} is illustrated by the exact results for $\alpha=0$ in Fig.~\ref{fig:phase}~c) - beyond the length scale $L_0$ the averaged correlator $X^2$ either ceases to be monotonous (for $p<p^{SB}_c$) or decays more quickly than the polynomial ansatz would predict (for $p>p^{SB}_c$). In consequence, the values of $X^2_{\infty}$ are smoothed in the vicinity of the transition point.

Moreover, we observe that the value of the extrapolated correlation function $X^2_{\infty}$ is reduced when the range of interactions is decreased (i.e. when $\alpha$ increases), as shown in Fig.\ref{fig:phase}~e). At the same time the deviations with respect to the mean field results are enhanced. Finally, for sufficiently short-range interactions (e.g. $\alpha=2$) we no longer observe the ordered phase and the system always remains in the disordered, paramagnetic phase. This illustrates the crucial impact of the range of interactions $\alpha$ on the symmetry breaking transition. At $\alpha=2$ our model behaves similarly to the short-range model investigated in \cite{Rossini20} in which there is no stable long-range order.

Our results so far indicate that our system hosts the long-range order for sufficiently small values of $p$ and $h$. This is intuitively clear: a large resetting probability leads to a significant decrease of $X^2$ within a single cycle of the time evolution; on the other hand, for large fields ($h\gtrsim 1.5$) the value of $X^2$ increases and decreases back to a small value during the unitary evolution -- see Fig.\ref{fig:phase}~b).

\subsection{Locating the symmetry breaking transition}

\begin{figure}
\vspace{0.25cm}
	\includegraphics[width=1\columnwidth]{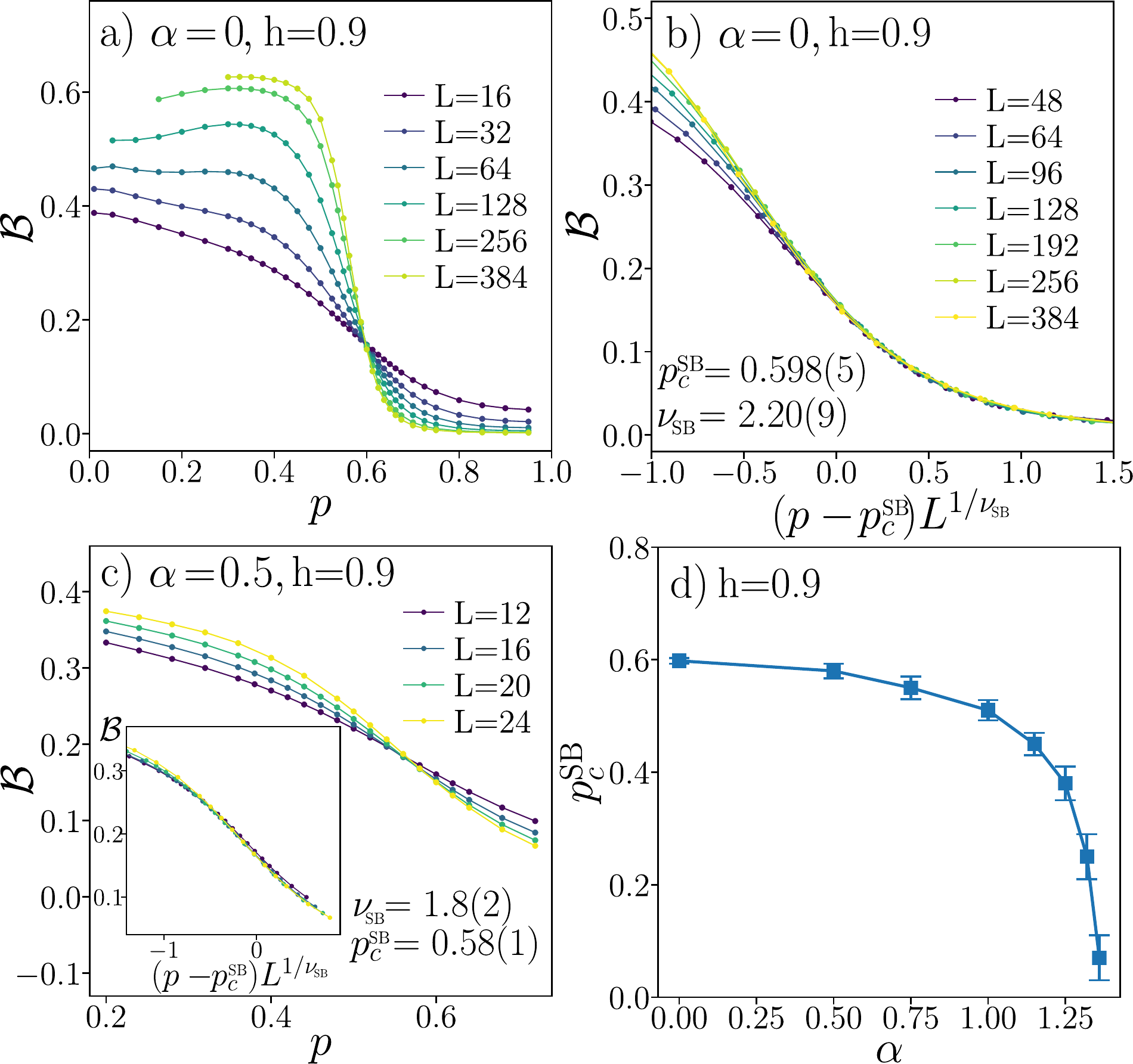}
	\caption{The Binder cumulant $\mathcal B$ across the symmetry breaking phase transition. Panel a) shows the Binder cumulant $\mathcal B$ as function of the resetting probability $p$ for various system sizes $L$ in the model with interactions of infinite range ($\alpha=0$). Panel b) shows the collapse of the Binder cumulant data with the scaling form \eqref{Eq:scalBind}. Panel c) shows Binder cumulant for system with $\alpha=0.5$,  the inset shows the collapse of the data using \eqref{Eq:scalBind}. Panel d) shows the dependence of the critical resetting probability $p^{SB}_c$ on the range of interactions $\alpha$.
  }
	\label{fig:bind}
\end{figure}

In this section we investigate the critical properties of the order-disorder transition in the kicked Ising model with resetting. In principle, the averaged two-point correlation function $X^2$ could be used as an order parameter for the symmetry breaking transition since it is larger than $0$ in the ordered phase and vanishes in the disordered phase. However, our results for large system sizes for the $\alpha=0$ case illustrate that $X^2$ is subject to strong finite-size effects: the asymptotic behavior starts only once the system size exceeds the length scale $L_0\approx 50$. Such system sizes are not accessible in our exact numerical simulations for $\alpha>0$. Hence, we consider the Binder cumulant \cite{Binder81, Binder81a}
\begin{equation}\label{Eq:Binder}
\mathcal B = 1-\frac{ x^4}{3 (x^2)^2}
\end{equation}
where $x^a=\overline{\ave{\psi(t)| (\sum_{i=1}^L \sigma^x_i)^a | \psi(t)}}$ for $a=2, 4$ and the overline denotes the average over quantum trajectories. The Binder cumulant is a function of the two-point and four-point correlation functions. For $\alpha=0$ we calculate the Binder cumulant as
\begin{equation}\label{Eq:Binder_rho}
\mathcal B = 1-\frac{ \text{Tr}\left(\rho_{ss}  (\sum_{i=1}^L \sigma^x_i)^4 \right) }{3 \left( \text{Tr}(\rho_{ss}  (\sum_{i=1}^L \sigma^x_i)^2) \right)^2},
\end{equation}
where $\rho_{ss}$ is the average steady state just after the resetting process. 

The Binder cumulant in the large system size limit vanishes in the paramagnetic phase and tends to $2/3$ in the ordered phase \cite{Binder81}. This trend is indeed followed by our results for the kicked Ising model with resetting at $\alpha=0$, as shown in Fig.~\ref{fig:bind}~a). The transition between the two phases sharpens up with increasing system size. We perform a finite size scaling analysis assuming a scaling form
\begin{equation}\label{Eq:scalBind}
\mathcal B = f[(p-p^{SB}_c) L^{1/\nu_{SB}}]
\end{equation}
where $f$ is a certain universal function, $p^{SB}_c$ is the critical resetting probability and $\nu_{SB}$ is a critical exponent. The obtained data collapse for $\alpha=0$ is shown in Fig.~\ref{fig:bind}~b). Its quality is very good in the $-0.5<(p-p^{SB}_c)L^{1/\nu_{SB}}<1.5$ interval. The deviations for $(p-p^{SB}_c)L^{1/\nu_{SB}}<-0.5$ are visible only for smaller system sizes ($L\leq128$) and could be taken into account by subleading corrections to the scaling form \eqref{Eq:scalBind}. The large interval of available system sizes allows for an accurate determination of the critical parameters: $p^{SB}_c=0.598(5)$ and $\nu_{SB}=2.20(9)$. 
We note that the discrepancy between resetting probability $p^{SB}_c$ (as determined by a finite size scaling of the results for $\alpha=0$) and the value obtained in the mean field approximation $p^{SB}_{MF}\approx 0.5819$, (obtained from the solution of Eq. \eqref{Eq:MFlinpc}) is small.
Importantly, when we restrict data for $\alpha=0$ to system sizes $L=12, ..., 24$, we observe that the scaling form \eqref{Eq:scalBind} can be used to obtain a data collapse in the restricted interval of system sizes yielding $\tilde{p}^{SB}_c=0.590(15)$ and $\tilde{\nu}_{SB}=1.8(2)$ which agree (within $2$ standard deviations in the case of  ${\nu}_{SB}$) with the critical parameters obtained for much larger system sizes. 

The last observation suggests that data for systems of size $L\leq24$, available for $\alpha>0$, can be analyzed with the scaling form \eqref{Eq:scalBind} yielding reasonable estimates of critical parameters. The results for $\alpha=0.5$, along with the obtained collapse, are shown in Fig.~\ref{fig:bind}~c). The curves $\mathcal B(p)$ cross around $p\approx0.58$ which is consistent with the value $p^{SB}_c=0.58(1)$ obtained in the finite size scaling according to \eqref{Eq:scalBind}. 

We repeat the scaling analysis of the Binder cumulant for other values of $\alpha>0$. The resulting values of the critical resetting probability $p^{SB}_c$ are shown in Fig.~\ref{fig:bind}~d). The crucial impact of the range of interactions $\alpha$ on the symmetry breaking transition is clearly visible: the critical resetting probability diminishes with increasing $\alpha$. The values of the critical exponent are, within error bars equal to $\nu_{SB}\approx2$ for each considered value of $\alpha$. Finally, for sufficiently short-range interactions, $\alpha>\alpha_c$,  the critical resetting probability tends to $0$ and the system shows no symmetry breaking, always remaining in the disordered phase. Our numerical simulations suggest that this happens at $\alpha_c \approx 1.3$. However, this value of $\alpha_c$ may be underestimated due to large error bars on the values of $p^{SB}_c$ around $\alpha=1.25$.

\begin{figure}
	\includegraphics[width=1\columnwidth]{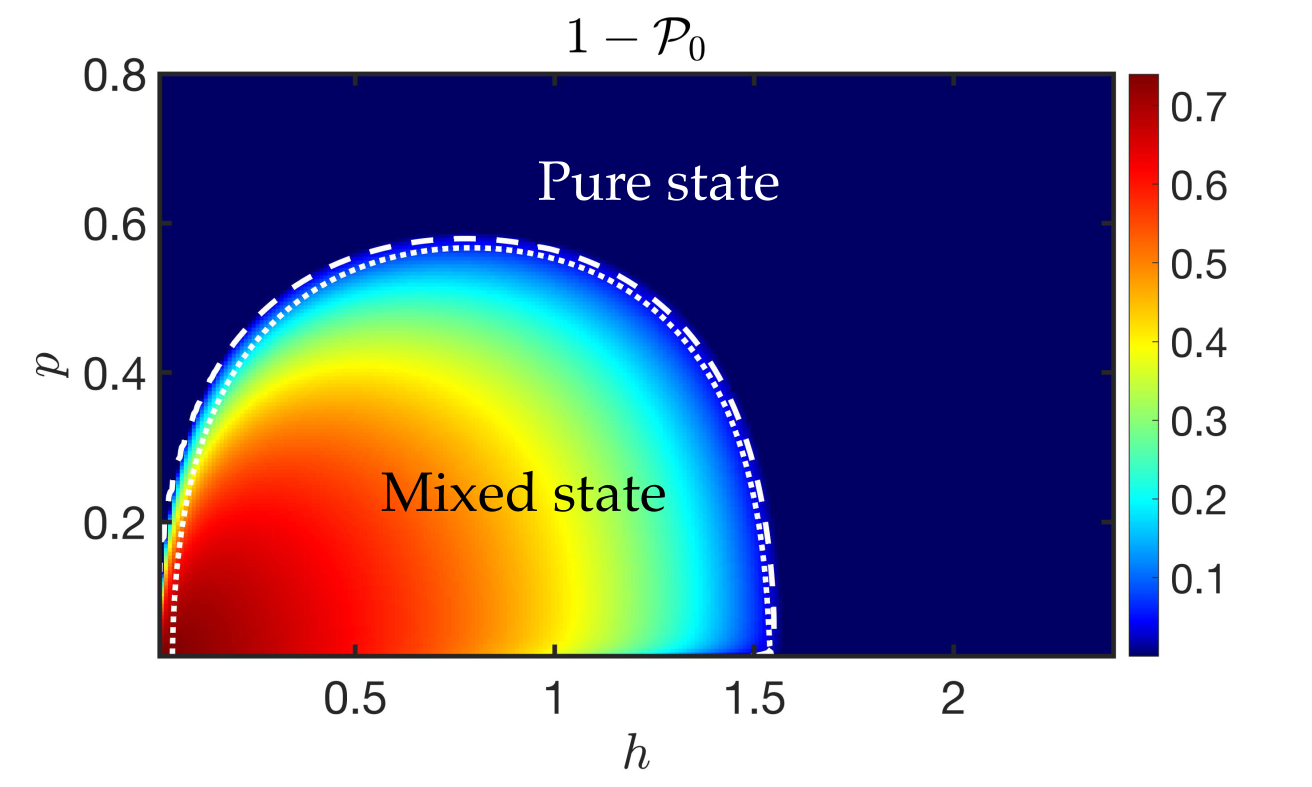}
	\caption{{Transition in the purity of the average state in the mean field approximation. The plot shows a color map of $1-\mathcal P_{0}$ as function of $h$ and $p$. The white dashed line shows the contour line $1-\mathcal P_0=0.01$, while the dotted line shows the contour of the magnetization $|\overline{\langle\sigma^x\rangle}|=0.01$ calculated from mean field. }}
	\label{fig:Color_map_purity}
\end{figure}

\subsection{Purity of the average state}\label{Sec:orderPurity}

In this Section we analyze the purity of the average state, i.e. we calculate the purity of the system after performing the average over the quantum measurements outcomes. To that end, we analytically study
\begin{equation}\label{Eq:PurAVG}
\mathcal{P}_{\mathrm{avg}}\equiv\text{Tr}\rho_{\mathrm{ss}}^2=\text{Tr}(\overline{\ket{\psi}\bra{\psi}}\cdot\overline{\ket{\psi}\bra{\psi}}),
\end{equation}
within mean field approximation and then compare the results with exact numerical simulations of the full density matrix of the system. The system is initialized in a fully mixed state and we investigate whether the steady-state is mixed or pure. 

\emph{Mean field} -- In order to study $\mathcal{P}_{\mathrm{avg}}$ we employ a mean field analysis (see Section \ref{Sec:methodsMF}).

Solving the mean field equations for the dynamics of the system, we compute the steady state value of $\mathcal P_{\mathrm{avg}}$ as $\mathcal P_{\mathrm{avg}}=(\mathcal P_0)^L$, where $\mathcal P_0=(1+|\vec s|^2)/2$ is the purity per site and $\vec s=\mathrm{Tr}(\vec \sigma_i \rho)$ is the local magnetization.
Our results are presented in Fig. \ref{fig:Color_map_purity}. At the level of mean field approximation the symmetry breaking transition and the transition between mixed and pure steady state coincide.

\begin{figure}
\vspace{0.25cm}
	\includegraphics[width=1\columnwidth]{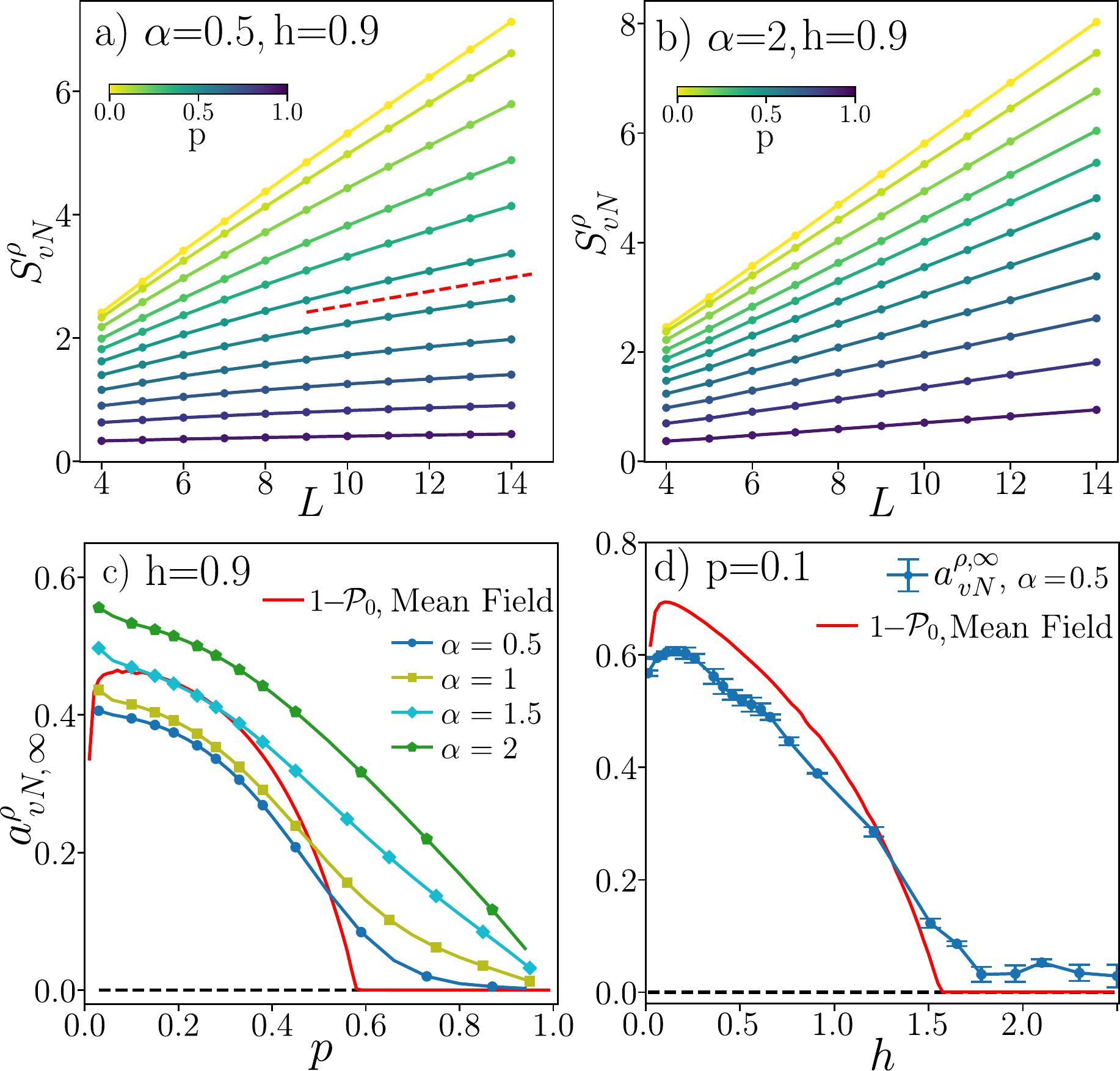}
	\caption{Purity and entropy of the steady state $\rho_{ss}$. Panels a) and b) show the von Neumann entropy of the steady state density matrix $S^{\rho}_{vN}$ as a function of system size for various resetting probabilities $p$, the red dashed line in panel a) separates the data for $p>p^{SB}_{c}$ and $p<p^{SB}_{c}$. Panels c) and d) show the extrapolated coefficient governing the system size scaling of $S^{\rho}_{vN}$ together with $1-\mathcal{P}_0$, where $\mathcal P_{0}$ is the purity (per site) of the steady-state obtained in the mean field approximation.
  }
	\label{fig:purification}
\end{figure}

\emph{Exact time evolution of the density matrix} -- We now turn to exact numerical simulations of the evolution of the full density matrix $\rho$. We initialize the system in the completely mixed state $\rho=I/2^L$ and calculate $\rho(t)$. To that end, we perform the full exact diagonalization of the evolution operator \eqref{Eq:IsHam}, which allows us to calculate $\rho(t+T)=\mathcal{U} \rho(t) \mathcal{U}^{\dag}$. Each step of the unitary time evolution is followed by the resetting $\rho'(t)=\prod_{i=1}^L \sum_{\mu_i=0}^2 K_{i,\mu_i} \rho(t) K^{\dag}_{i,\mu_i}$. Calculations with full density matrices are computationally much more demanding than keeping track of individual trajectories of the system, therefore results presented in this section are for system sizes $L\leq14$. 

We calculate the von Neumann entropy of the density matrix 
\begin{equation}\label{Eq:SVN}
S^{\rho}_{vN}(t)\equiv- \text{Tr}\left[ \rho(t) \ln(\rho(t) ) \right]
\end{equation}
at time $t=5L$. 
The results are shown in Fig.~\ref{fig:purification}~a),~b) as function of system size $L$. 
We observe two different scaling regimes of the
entropy density $S^{\rho}_{vN}/L$: it is constant in the 'mixed' regime and vanishes in the 'pure' regime. Indeed, at small resetting probabilities $S^{\rho}_{vN}$ grows linearly with system size $L$ and this growth is much slower at large values of $p$. 

We take the following steps to quantitatively examine the two scaling regimes of the entropy of the average state. First, we locally (at system sizes $L-1, L, L+1$) fit $S^{\rho}_{vN}$ with a linear function: $S^{\rho}_{vN}(L) = aL+b$. Performing the fitting at various $L$ we obtain the coefficient $a(L)$ which we then fit with a second order polynomial in $1/L$: $a(L)=a^{\infty}_{vN}+b_1/L+b_2/L^2$. The value of $a^{\infty}_{vN}$ is our estimate of the asymptotic value of coefficient $a(L)$ determining the growth of $S^{\rho}_{vN}$ with system size. The mixed regime is characterized by $a^{\infty}_{vN}>0$ and hence by a constant entropy density $S^{\rho}_{vN}/L$, whereas the pure regime is found when $a^{\infty}_{vN}=0$. 

The results of this analysis, along with $1-\mathcal P_{0}$ are shown in Fig.~\ref{fig:purification}~c),~d). 
The regime in which $a^{\infty}_{vN}$ is non-zero, obtained from the numerical simulations of our model at $\alpha=0.5$, agrees well with the interval of parameters for which mean field predicts a non-zero value of $1-\mathcal{P}_0$. This applies to data for fixed $h=0.9$ as a function of the resetting probability $p$, as well to results for fixed $p=0.1$ as a function of the field $h$. Comparison of the results of the mean field approach with the behavior of $a^{\infty}_{vN}$ suggests that the latter might be smoothed out around the transition point due to residual finite size effects that are present in the numerical data. Our numerical results do not allow us to distinguish whether the change in scaling of the entropy of the steady state becomes a transition in the thermodynamic limit or is a smooth crossover. However, we clearly observe that when the range of interactions is decreased ($\alpha$ is increasing), the regime in which the state of the system remains mixed extends to larger values of the resetting probability $p$. Finally, for $\alpha=2$, there is no pure regime of the average state and the system remains always in the mixed state. The asymptotic rate of increase of entropy of the steady state $a^{\infty}_{vN}$ stays positive for all values of $p$ for $\alpha=2$ (see Fig.~\ref{fig:purification}~c) even though it is a decreasing function of $p$ which shows that more frequent resetting diminishes the number of states available to dynamics of the system. 

Investigations of symmetry breaking transition and of the (putative) transition in the purity of the average state are two complementary ways of characterizing the steady state of the system $\rho_{ss}$. The two transitions share the same dependence on the resetting probability $p$ and on the field $h$ for sufficiently long-range interactions ($\alpha \lesssim 0.5$) and coincide at the level of mean field approximations. However, the two transitions are affected by the decrease of the range of interactions in an opposite manner. At $\alpha=0.5$, the regime of parameters in which the steady state is mixed corresponds to the ordered phase. With a decrease in the range of interactions (increase of $\alpha$) the ordered phase shrinks to smaller values of the resetting probability $p$ whereas the regime of the mixed steady state extends to larger and larger values of $p$. Finally, for sufficiently short-range interactions (e.g. $\alpha=2$), the system hosts no long-range order and the steady state always remains mixed.

\section{Entanglement transition}\label{Sec:entanglement}

The crucial feature of the ordering transition is an abrupt change in the properties of the average state of the system $\rho$. In contrast, the entanglement transition occurs at the level of individual quantum trajectories: it separates phases with volume-law and area-law (or sub-volume law for systems with power-law interactions, see \cite{Block2021}) scaling of the entanglement entropy of the state propagated along the quantum trajectory. The entanglement transitions were firstly described in quantum circuits with projective measurements \cite{Li18, Li2019,  Skinner2019} and arise due to the competition between unitary dynamics, that tends to increase the entanglement, and local measurements, which suppress the long-range entanglement in the system. Entanglement transitions were found also in many-body systems undergoing Hamiltonian evolution with random measurements \cite{Tang20, Goto20, Rossini20, Lunt20}. From the perspective of protection of quantum information against the non-unitary evolution of the system, the volume-law phase can be identified as a quantum error-correcting phase in which initially mixed state gets purified at time scale exponential in system size, whereas in the area-law phase the purification time is system size independent \cite{Gullans2020}. Importantly, this distinction between the error correcting and the quantum Zeno phases generalizes to systems with long-range interactions for which the notion of area-law phase may be ill-defined. 

Entanglement phase transitions may be probed via suitable unconventional correlation functions \cite{Buchhold2021,Lu21} or via a non-linear functional of the density matrix, such as the entanglement entropy or the negativity. The entanglement entropy is perhaps the most natural probe, since in 1D systems with local interactions it scales linearly with size in the volume-law phase and is system size independent in the area-law phase. However, the logarithmic scaling of the entanglement entropy at the transition between volume-law and area-law phases impedes a precise characterization of the transition even in the numerically tractable case of stabilizer circuits.

Instead, bipartite quantum mutual information (QMI) \cite{Li2019, Szyniszewski19} and tripartite QMI \cite{Zabalo20} were proposed as more reliable tools to pinpoint the transition. To investigate measurement induced phase transition in the Kicked Ising model with resetting we proceed as follows. In the next subsection we investigate the entanglement phase transition using the QMI and, in the following subsection, we look at the transition directly through the lens of entanglement entropy. In the subsequent subsection, we show that the volume-law and sub-volume law phases determined by this approach correspond respectively to quantum error correcting and quantum Zeno phases.

\subsection{Quantum mutual information}
\label{Subsec:QMI}

\begin{figure}[!t]
\vspace{0.25cm}
	\includegraphics[width=1\columnwidth]{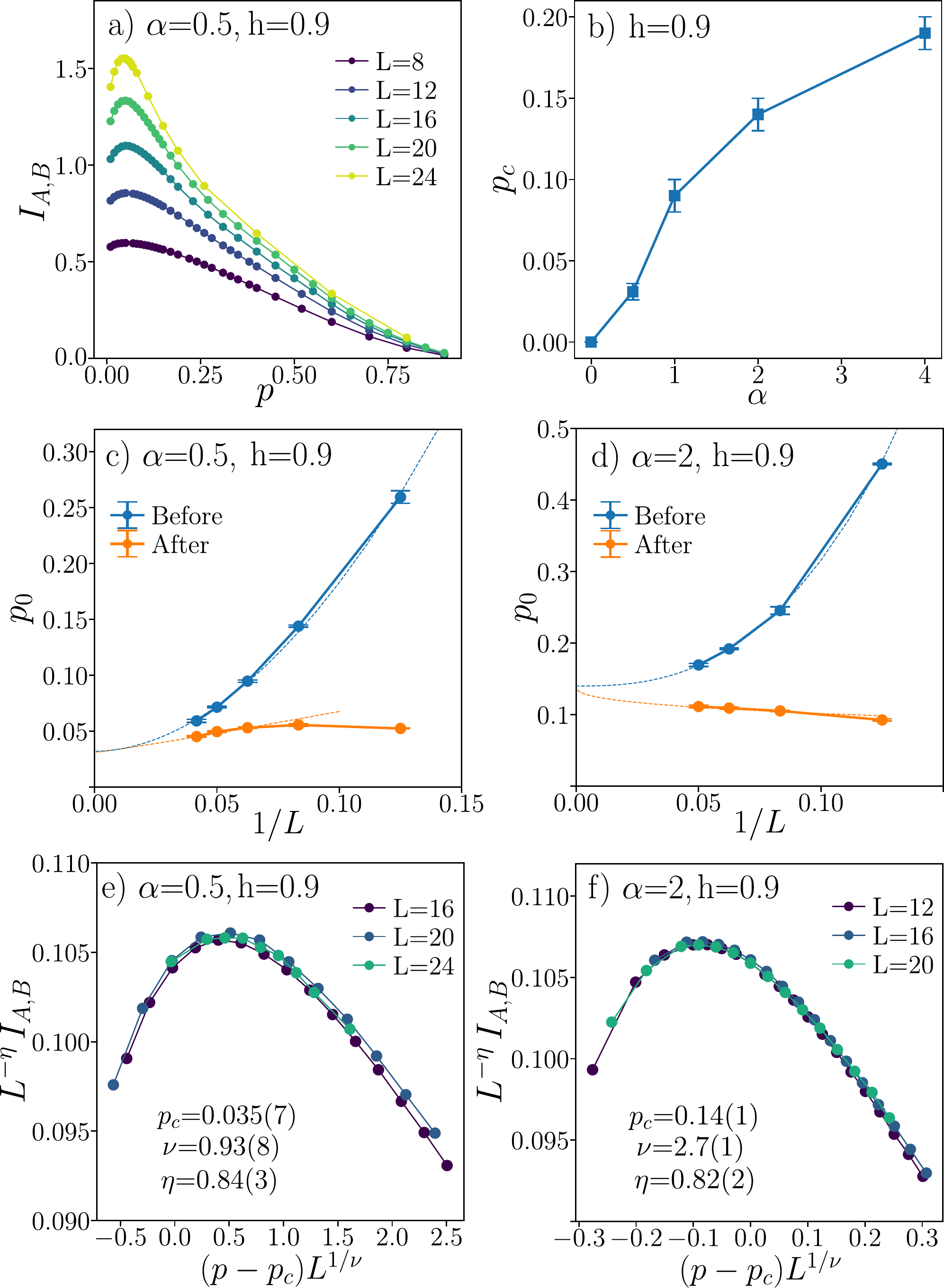}
	\caption{The entanglement phase transition. Panel a) shows the quantum mutual information $I_{A,B}$ where $A, B$ are subsystems of size $L/4$ located adjacently in the system of size $L$ with periodic boundary conditions, calculated after the resetting. Panel b) shows the critical measurement probability $p_c$ as a function of the range of interactions $\alpha$. Panels c) and d) show the measurement probability $p_0$ corresponding to the maximum of $I_{A,B}$ for various system sizes respectively for models with $\alpha=0.5$ and $\alpha=2$, the curves denoted by 'After' ('Before') correspond to measurement of $I_{A,B}$ before (after) the resetting. Panels e) and f) show the collapse of $I_{A,B}$ data onto universal curves for $\alpha=0.5$ and $\alpha=2$ respectively.
  }
	\label{fig:qmi}
\end{figure}

To investigate the properties of the entanglement transition in our model we employ the quantum mutual information, defined as
\begin{equation}\label{Eq:qmi}
I_{A,B} = S(A)+S(B)-S(A\cup B),
\end{equation}
where $A$ and $B$ are subsystems of size $|A|=|B|=L/4$ located adjacently on the chain with periodic boundary conditions, and $S(A)= -\text{Tr}\left[ \rho_A \ln(\rho_A ) \right]$, with $\rho_A$ being the reduced density matrix of subsystem $A$ in state $\ket \psi$.

The QMI measures the total amount of correlations between the subsystems $A$ and $B$ \cite{Groisman05}. In the area-law phase, the state is 'close' to a product state, hence one may expect weak correlations between the two subsystems and, consequently, a small value of QMI. Well into the volume-law phase, the subsystems $A$ and $B$ are strongly entangled with their exterior but the amount of information shared between $A$ and $B$ is small. Thus, one may expect $S(A)+S(B)$ to be close to $S(A \cup B)$, and the QMI to be small in the volume law phase as well. On the other hand, at the entanglement phase transition the correlations between subsystems $A$ and $B$ are enhanced and the QMI is maximal \cite{Skinner2019}. While the above description applies directly to systems with local couplings, we show below that the the behavior of our system with power-law interactions is fully analogous. 
 
Fig.~\ref{fig:qmi}~a) shows the behavior of the QMI as a function of resetting probability $p$, for a system with $\alpha=0.5$. For a given system size $L$, the curve $I_{A,B}(p)$ has a well defined maximum $p_0(L)$. Since the maximum sharpens up with increasing system size, it is plausible to assume that $p_0(L)$ approaches the value of the critical resetting probability $p_c$ in the thermodynamic limit $L\rightarrow \infty$, similarly to stabilizer circuits \cite{Skinner2019}. More in detail, $p_0(L)$ is extracted from the QMI measured just before (after) the resetting of the spins. The difference between $p_0$ obtained from the QMI measured before and after the resetting is quite large for small system size $L\approx 10$, as Fig.~\ref{fig:qmi}~c),~d) illustrates, but this difference diminishes with increasing system size and the two curves seem to converge to a single critical resetting probability $p_c$.

The resulting values of $p_c$ versus $\alpha$ are shown in Fig.~\ref{fig:qmi}~b). We observe that in the limit of  all-to-all coupling, $\alpha \rightarrow 0$, the transition point approaches $p_c\rightarrow 0$, implying that for $\alpha=0$ the system is always in a sub-volume law phase. As the range of interactions is decreased, the entanglement phase transition shifts to a finite value of resetting probability. While close to $\alpha=0$ $p_c$ is sharply increasing with $\alpha$, this is no longer the case beyond $\alpha=2$. Nevertheless, in the limit of strictly local interactions ($\alpha \rightarrow \infty $), we expect a non-zero value of $p_c$ \cite{Li2019}. Remarkably, the critical resetting probability at all considered interaction ranges obeys the bound $p_c<0.1893$ of \cite{Fan21} derived under the assumption that the volume-law state is an encoding of a Page state in a quantum error-correcting code.
 
To quantitatively characterize the transition, we perform a finite-size scaling analysis of QMI using the scaling form
\begin{equation}\label{Eq:QmiScaling}
 I_{A,B}(p) =L^{\eta}f[ (p-p_c)L^{1/\nu} ],
\end{equation}
where $f(x)$ is assumed to be a certain universal function, $p_c$ is the critical probability of resetting, $\eta$ and $\nu$ are critical exponents. For stabilizer circuits with short-range interactions \cite{Li2019} the exponent $\eta=0$, i.e. the maximum of QMI saturates to a constant that is independent of the system size $L$. This is a manifestation of an emergent conformal symmetry at the transition, that arises also in 1D quantum critical systems at equilibrium \cite{Calabrese09}. The conformal symmetry is broken when long-range interactions are introduced to the stabilizer circuits \cite{Block2021,Sharma21}. Then, the maximum of QMI increases with system size and $\eta>0$.

The obtained collapses of QMI data are shown in Fig.~\ref{fig:qmi}~e),~f) respectively for models with $\alpha=0.5$ and $\alpha=2$. In the finite size scaling analysis we considered only the QMI measured after the resetting process because it is characterized by a weaker system-size dependence compared to the QMI measured before the resetting. Hence, we assume that the former reflects the asymptotic properties of the system more reliably. For both $\alpha=0.5$ and $\alpha=2$, the maximum of $I_{A,B}(p)$ increases nearly linearly with system size $L$, and the obtained values $\eta=0.84(3)$ ($\eta=0.82(2)$) are close to unity. Moreover, we observe a decrease of the critical resetting probability (from  $p_c=0.14(1)$ to $p_c=0.035(7)$) and of the value of the exponent $\nu$ (from $\nu=2.7(1)$ to $\nu=0.93(8)$) as the range of interactions is increased from $\alpha=2$ to $\alpha=0.5$. 

The behavior of the critical resetting probability $p_c$ and of the critical exponent $\nu$ as function of the interaction range is different from what has been observed for long-range stabilizer circuits \cite{Block2021,Sharma21} and for systems of spinless fermions with long-range interactions \cite{Minato21}. A direct comparison between our results is impossible because of the fundamental qualitative differences in our models. However, we can point out some of the possible reasons behind the different observations. For example, we adopt a Kac prescription specified in Eq. \eqref{Eq:KacN}, while the work  \cite{Minato21} does not take into account such terms making the interactions they consider more effective at enhancing entanglement growth. Moreover, in \cite{Block2021} the authors consider Clifford gates with range drawn from a power-law distribution. The number of pairs of interacting sites is always of order $L$, while our model may exhibit a larger number of interaction pairs (up to $\sim L^2$ in the $\alpha\rightarrow0$ limit because of the all-to-all interactions) but weaker compared to \cite{Block2021} because of the Kac renormalization. This significant difference is likely the primary source of the different entanglement behavior.

The behavior of the system across the entanglement transition is similar when the QMI \eqref{Eq:qmi} is calculated using the Renyi entropy $S(A) \rightarrow S_n(A)=\left[\ln\text{Tr} \rho_A^n \right]/(1-n)$ with $n \geq 2$ instead of the von Neumann entanglement entropy. Performing a scaling procedure analogous to the one presented in Fig.~\ref{fig:qmi}~e),~f), we find $p^{n=2}_c=0.051(5)$, $\nu=0.91(5)$, $\eta=0.75(1)$ for $\alpha=0.5$ and $p^{n=2}_c=0.19(2)$, $\nu=2.70(5)$, $\eta=0.72(1)$ for $\alpha=2$. The discrepancies between the results for von Neumann and Renyi entanglement entropies likely stem from finite size effects which are expected to be stronger for Renyi entropy.

We note that the subsystems $A$ and $B$ of size $|A|=|B|=L/4$ considered  in the calculation of QMI \eqref{Eq:qmi} are located adjacently. In contrast, the subsystems $A'$ and $B'$ located on \emph{antipodes} of the ring with periodic boundary conditions were used to pin-point the MIPT in \cite{Li2019}. We have checked that the latter configuration gives rise to quantitatively the same results as the former provided that $\alpha \lesssim 1$. This is intuitively expected: for sufficiently small $\alpha$, the notion of locality in the system is lost and the spatial configuration of the of subsystems is not relevant. For shorter range interactions, e.g. at $\alpha=2$, the configuration with subsystems on the antipodes yields inconclusive results about MIPT in our system: in particular, it seems to be strongly affected by finite volume effects. This suggests that the choice of neighboring subsystems allows for a better location of MIPT in the systems with non-local interactions. The insights into MIPT in the Kicked Ising model with resetting obtained from QMI are further confirmed by system size scaling of bipartite entanglement entropy.

\subsection{Entanglement entropy at the entanglement phase transition}\label{subsec:Ententa}

\begin{figure}
\vspace{0.25cm}
	\includegraphics[width=1\columnwidth]{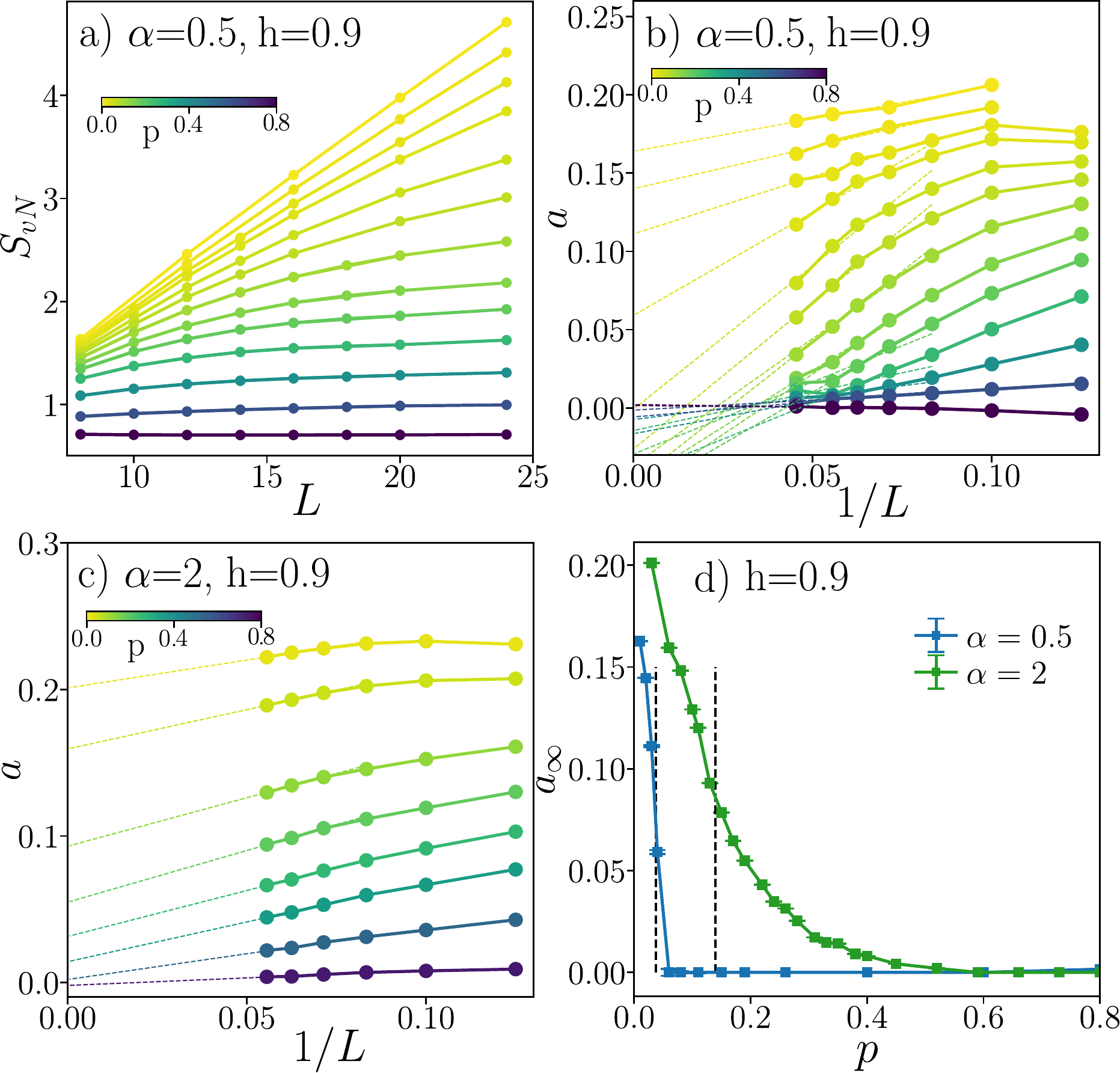}
	\caption{The bipartite entanglement entropy $S_{vN}$ at the entanglement phase transition. Panel a) shows $S_{vN}$ in steady state of system of size $L$, for $\alpha=0.5$ and $h=0.9$. Panels b) and c) show the slope $a$ of the $S_{vN}(L)$ curve as a function of system size $L$ for various resetting probabilities and $\alpha=0.5, 2$. Panel d) shows the coefficient $a_{\infty}$ obtained from extrapolation of the slope $a$ to thermodynamic limit $L \rightarrow \infty$. Unphysical negative values are replaced with $a_{\infty}=0$.
  }
	\label{fig:ent}
\end{figure}

In this subsection we consider the von Neumann bipartite entanglement entropy $S(A)= -\text{Tr}\left[ \rho_A \ln(\rho_A ) \right]\equiv S_{vN}$ (where the subsystem $A$ constitutes a half of the chain), and investigate the behavior of $S_{vN}$ across the entanglement phase transition. The entropy $S_{vN}$ is plotted as a function of system size $L$ for various resetting probabilities in Fig.~\ref{fig:ent}~a). A clear difference is visible between the regime of small resetting probabilities $p$ where $S_{vN}$ is increasing linearly with $L$ and large $p$ where $S_{vN}$ is independent of the system size.

To analyze the scaling of $S_{vN}$ more quantitatively, we follow a route similar to Sec.~\ref{Sec:orderPurity}: we fit $S_{vN}$ with a linear function: $S_{vN}(L) = aL+b$ using points at system sizes $L, L+2, L+4$. Changing the value of $L$, we obtain the coefficient $a(L)$ which we then fit with a first order polynomial in $1/L$: $a(L)=a_{\infty}+b_1/L$, using $a(L)$ for three largest system sizes available. The values of the coefficient $a$ together with the fits are shown in Fig.~\ref{fig:ent}~b),~c).

For $\alpha=0.5$, a positive value of $a_{\infty}$ is obtained only for $p<0.04$, as shown in Fig.~\ref{fig:ent}~d). For larger resetting probability, the coefficient $a$ decreases abruptly with system size which could suggest a negative value of $a_{\infty}$. Such a solution is unphysical since it implies a decrease of $S_{vN}$ with increasing system size. A plausible explanation is that a fit up to the linear order in $1/L$ could be a good approximation for $a(L)$ only for system sizes larger than the ones we have access to, and that it is not sufficient to fully capture the behavior of $S_{vN}$ for the smaller values of $L$ that we consider. Using a crude approximation, whose consequences we will scrutinize below, whenever the fit yields a negative value of $a_{\infty}$, we assume that the actual value is $a_{\infty}=0$, consistent with an area-law phase.

The problems with the extrapolation of our results to the thermodynamic limit for $\alpha=0.5$ illustrate that our method gradually looses its accuracy as the range of interactions becomes sufficiently large. Hence, even though the maximum of the QMI $p_0\rightarrow 0$ for $\alpha \rightarrow 0$, as shown in Fig.~\ref{fig:qmi}~b), we cannot exclude the emergence of a sub-volume law phase extending for $p>0$ at sufficiently small $\alpha$. In such a phase, the entanglement entropy could scale sub-linearly with system size, for instance as $S_{vN}\sim L^{\beta}$ with $0<\beta<1$. The existence of such a phase is appealing due to the extremely long-range of interactions at small $\alpha$, but to demonstrate or exclude such a possibility one would have to repeat our analysis at much larger system sizes. We note that the method exploiting the permutation symmetry on the system at $\alpha=0$ from Sec. \ref{Sec:methods_SYMM} can only be used to calculate the average state $\rho_{ss}(t)$, which does not contain information about entanglement properties of states at individual quantum trajectories, and cannot be used to shed light on the existence of a sub-volume law phase at $\alpha=0$. 

For systems with shorter range interactions ($\alpha=2$), we do not encounter problems with the extrapolation of the behavior of von Neumann entanglement entropy and find that $a_{\infty}$ is a monotonous function of $p$ decreasing to $0$ at sufficiently large $p$. The resulting values of $a_{\infty}$ are shown in Fig.~\ref{fig:ent}~d). The critical resetting probabilities $p_c$ obtained from the finite-size analysis of the QMI, correspond to regions where the value of $a_{\infty}$ quickly decreases with $p$, showing a high degree of consistency between the two approaches. The residual, non-zero value of $a_{\infty}$ at $p>p_c$ shows that there are finite-size effects in the system that cannot be fully grasped by our scaling analysis.

\subsection{Purification transition}
\label{sec:app_puri}

\begin{figure}[!t]
\vspace{0.25cm}
	\includegraphics[width=1\columnwidth]{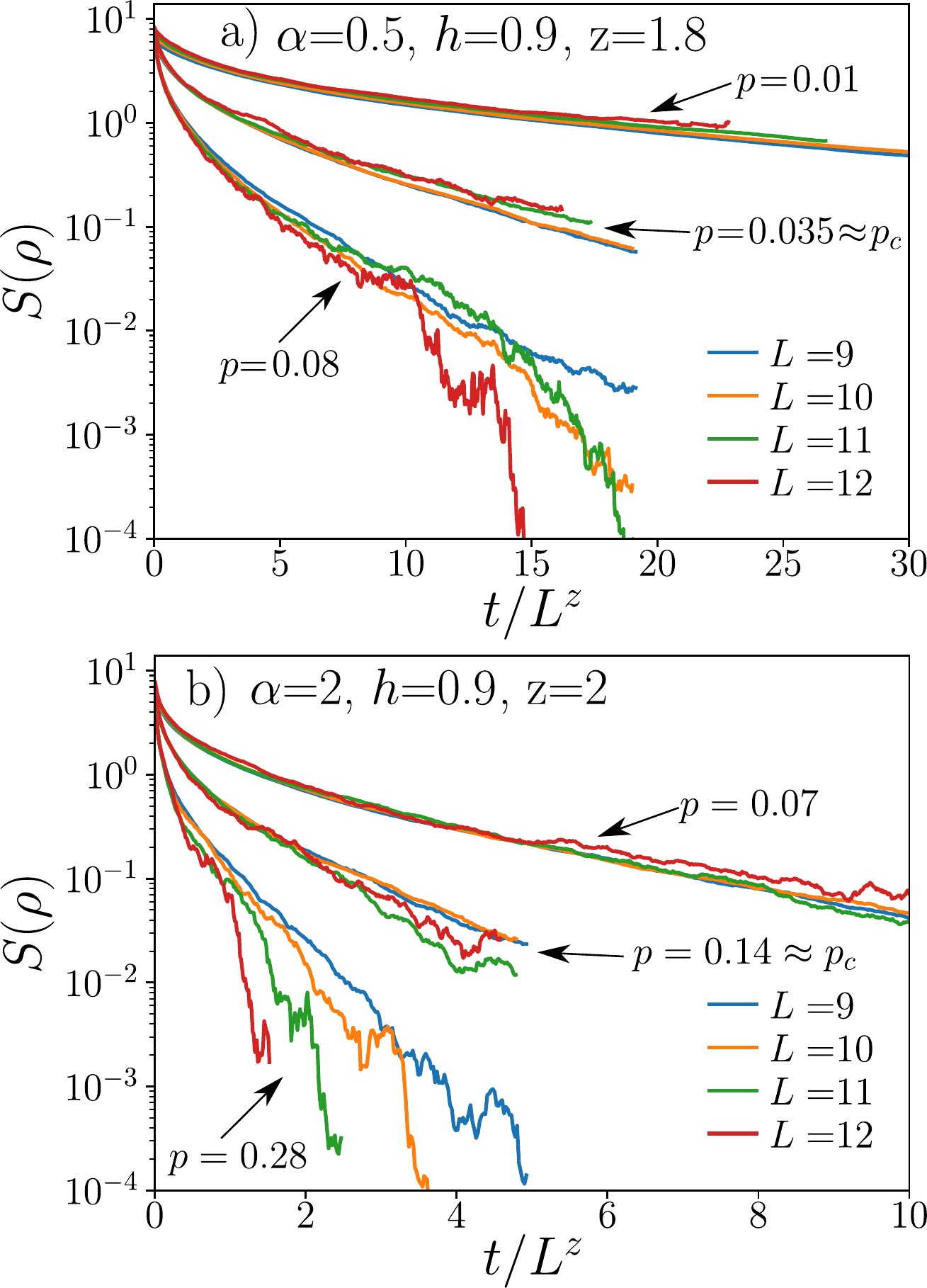}
	\caption{ Purification phase transition. The average von Neuman entropy of the density matrix $S(\rho)$ is plotted as function of time $t$ rescaled by $L^z$, where $L$ is the system size and $z$ is the dynamical critical exponent. Panels a) and b) show results for systems with $\alpha=0.5$ and $\alpha=2$. For $p<p_c$, the value of $S(\rho)$ at (sufficiently large) fixed $t/L^z$ increases as data for $p=0.01$ ($p=0.07$) demonstrate for $\alpha=0.5$ ($\alpha=2$), the opposite trend is visible for $p>p_c$ -- as visible from data for $p=0.08$ at $\alpha=0.5$ and for $p=0.28$ at $\alpha=2$  At the transition, $p\approx p_c$,  $S(\rho)$ is approximately constant at fixed value of $t/L^z$.
  }
	\label{fig:purif}
\end{figure}

So far, measurement induced criticality has been primarily characterized as an entanglement transition: a pure state is evolved under measurements and unitary dynamics and its bipartite entanglement entropy is monitored. Another facet of measurement induced criticality has been proposed in Ref. \cite{Gullans2020} and corresponds to {\it purification transitions}. In this context, the system -- which again is subject to a combination of measurements and unitary dynamics -- is instead initialized in a mixed state. 
By monitoring outcomes of measurements, one continually gains information about the system \cite{Zoller87}, so that the number of states consistent with the measurement record is steadily decreasing. This reduces the entropy of the state $\rho^{(c)}$ of the system, and may lead to its purification. However, the chaotic unitary evolution of the quantum many-body system counteracts this process by scrambling the quantum information and protecting it against local measurements \cite{Choi20}. This competition leads to a dynamical phase transition between a "mixed" phase in which the system purifies at time scale exponential in system size, and a "pure" phase in which the state purifies at a constant system-size-independent rate.

To investigate the presence of a purification transition in the spin chain of our interest, we consider the conditioned density matrix $\rho^{(c)}$ along a trajectory, initialize it in a fully mixed state $\rho^{(c)}=I/2^L$ and calculate its time evolution. 
To observe the purification phase transition, after each step of the unitary time evolution, we transform the density matrix according to 
\begin{equation}
    \rho^{(c)} \rightarrow \frac{  K_{\mu}\rho^{(c)} K_{\mu}^{\dag} }{ \mathrm{Tr}[K_{\mu}\rho^{(c)} K_{\mu}^{\dag}]}
\end{equation}
at each resetting step, with probability $\mathrm{Tr}[K_{\mu}\rho^{(c)} K_{\mu}^{\dag}]$ \cite{Li21}.

The entanglement transition manifests itself in a purification of the density matrix $\rho^{(c)}(t)$. 
Note that in section \ref{Sec:orderPurity} we were interested in properties of the average steady state $\rho=\overline{\rho^{(c)}}$, for which the unitary evolution $\rho(t+T)=\mathcal{U} \rho(t) \mathcal{U}^{\dag}$ was followed by the resetting $\rho(t) \rightarrow  \sum_{\mu} K_{\mu} \rho(t) K^{\dag}_{\mu}$. Such a time evolution results after sufficiently long time in the steady state $\rho_{ss}$. In this section, we are instead interested in following the evolution of single trajectories, in a similar way as for the detection of the entanglement transition. 

The results of our numerical simulations are shown in Fig.~\ref{fig:purif}. For each trajectory of matrix $\rho^{(c)}$, we calculate its von Neumann entropy and then average the results over hundreds of trajectories, obtaining
\begin{equation}
\label{Srho}
S(\rho)=-\text{Tr}[\overline{\rho^{(c)} \ln \rho^{(c)}}]
\end{equation}

Due to the average over trajectories, the calculations are very time consuming, therefore we are limited to system sizes $L\leq12$. 

In the first step we determine the values of dynamical critical exponent $z$. The exponent $z$ is not equal to unity as is in the stabilizer circuits \cite{Gullans2020a}, due to lack of conformal invariance in our system. To calculate $z$ we use the values of $p_c$ for the entanglement phase transition estimated in Sections~\ref{Subsec:QMI} and \ref{subsec:Ententa}, and determine $z$ from the requirement that $S(\rho)$ for various system sizes $L$ collapse on universal curves at $p=p_c$. This yields $z=1.8\pm0.2$ ($z=2.0\pm0.2$) for $\alpha=0.5$ ($\alpha=2$). Those values of the dynamical critical exponent are close to the values found in Section~\ref{subsec:Expenta}, although their values are less certain due to the small interval of system sizes used in their determination.

Inspecting the behavior of $S(\rho)$ as function of $t/L^z$ for different system sizes in Fig.~\ref{fig:purif} we note that for $p<p_c$ the value of $S(\rho)$ slowly increases with $L$ for a fixed value of $t/L^z$. The opposite trend is visible for $p>p_c$, when $S(\rho)$ decreases with system size $L$ for a fixed value of $t/L^z$. Hence, for $p>p_c$ the state of the system is dynamically purified, exactly in the same manner as it was observed in the area-law phase of stabilizer circuits in \cite{Gullans2020a}.

\section{Multifractality in the steady state regime}\label{Sec:mulfrac_ss}

In this section we investigate the extent to which the wave function spreads over the Hilbert space corresponding to the ensemble of individual quantum trajectories. This allows us to provide a connection to the well-understood phenomenology of localization. The figure of merit we consider are participation entropies \cite{Mace19}
\begin{equation}\label{Eq:PartEnt}
 S_q =\frac{1}{1-q}\ln \left( \sum_{\beta=1}^{2^L} | \psi_{\beta}|^{2q}\right),
\end{equation}
where $| \psi_{\beta}|^{2q}$ is the $q$-th moment of the wave function expressed in a given basis of the Hilbert space. The asymptotic behavior of the participation entropies $S_q$ with the system size $L$ distinguishes wave functions that are delocalized, multifractal and localized. Hence, $S_q$ are of natural interest in studies of localization, both in single particle systems -- for instance in the 3D Anderson model \cite{Rodriguez10} -- and in many-body systems \cite{Beugeling15, Mace19, Luitz20}. For delocalized wave functions, $S_q= L \ln 2 $; oppositely, localized wave functions are characterized by a system size independent value of $S_q$, whereas for the multifractal wave functions  $S_q=D_q L$ where $D_q<\ln 2$ is a $q$-dependent multifractal dimension.
Before discussing the relevance of participation entropies for the physics we are interested in, it is worth noting that those may be of experimental relevance since the stochastic sampling of wave functions has become available in atomic as well as in solid state platforms~\cite{Brydges19,Leseleuc19,Chiaro20,Pagano2020quantum,Scholl20,Ebadi20,Zeiher17,Veit21}.

\subsection{Participation entropy in the Z basis}

\begin{figure}[!t]
\vspace{0.25cm}
	\includegraphics[width=1\columnwidth]{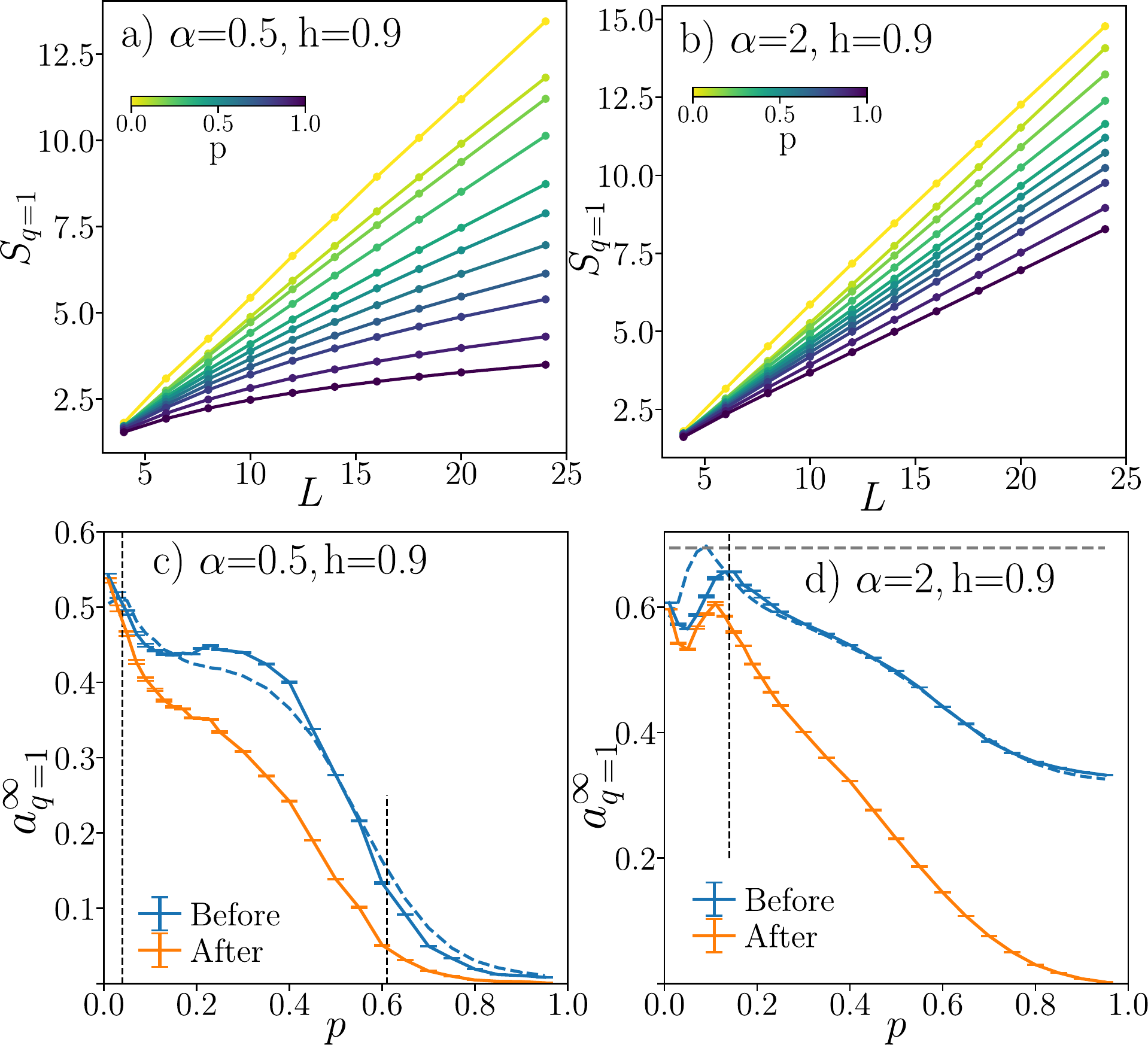}
	\caption{Participation entropy $S_{q=1}$ in the steady state of the system calculated in eigenbasis of $\sigma^z_i$ operators. Panels a) and b) contain the raw data for $S_{q=1}$ as a function of system size $L$ for the model with $\alpha=0.5$ and $\alpha=2$, respectively. Panels c) and d) show the coefficient $a^{\infty}_{q=1}$ in the system-size dependence of the participation entropy $S_{q=1}\sim a^{\infty}_{q=1} L $; the dashed blue lines correspond to extrapolations based on system sizes $L=8,...,14$, whereas the solid lines correspond to extrapolations based on $L=16,...,24$. Data denoted as 'Before' ('After') correspond to measurement of $S_q$ just before (after) the resetting. The vertical dashed lines show the position of the entanglement phase transition and symmetry breaking transition for $\alpha=0.5$ and the position of the entanglement phase transition for $\alpha=2$.
  }
	\label{fig:sq}
\end{figure}

One of the features of the participation entropy $S_q$ \eqref{Eq:PartEnt} is that it depends on the choice of basis in the Hilbert space. The eigenbasis of $\sigma^z_i$ operators (which we refer to as the 'Z basis') is a natural choice for the kicked Ising model with resetting. We expect that at large resetting probability the steady state will be close to the paramagnetic state $\ket{ \downarrow\downarrow ...\downarrow}$ and hence localized in the Z basis. Conversely, a decrease of the resetting probability may result in the gradual delocalization of the steady state.

Fig.~\ref{fig:sq}~a),~b) show the participation entropy $S_{q=1}$ as function of the system size $L$ for models with long-range ($\alpha=0.5$) and short range ($\alpha=2$) interactions. The participation entropy is calculated from the wave function before the resetting. For both $\alpha=0.5$ and $\alpha=2$, an increase in $p$ inhibits the growth of participation entropy with system size. 

In order to investigate the system size dependence of $S_q$ more quantitatively, we follow a strategy similar to the one employed for the bipartite entanglement entropy. We perform a linear fit $S_{q}(L) = a_q L+b$ using points at system sizes $L-2L, L, L+2$. This yields a coefficient $a_q(L)$. Subsequently, we extrapolate $a_q(L)$ to the $L\rightarrow \infty$ limit by fitting it with a second order polynomial $a^{\infty}_q+b_1/L+b_2/L^2$. The free coefficient of the polynomial $a^{\infty}_q$ is an estimate of the multifractal dimension $D_q$ that governs the scaling $S_q=D_q L$ of the participation entropy in the limit of large system size. 

The obtained values of $a^{\infty}_q$ (for $q=1$) are shown in Fig.~\ref{fig:sq}~c),~d). There are apparent differences in the coefficient $a^{\infty}_{q=1}$ obtained from the participation entropy measured before and after the resetting. Those differences are less pronounced in the volume-law phase at small resetting probability $p$. This resonates well with the point of view of quantum error correction: in the volume-law phase the local measurements are insufficient to alter the scaling of the participation entropy. Once the resetting probability surpasses the critical value for the entanglement phase transition, $p>p_c$, we observe a decrease in  $a^{\infty}_{q=1}$. Moreover, we note that there are apparent differences in the scaling of the participation entropy calculated before and after the resetting in the sub-volume law phase ($p>p_c$) reflected by differences in the respective $a^{\infty}_{q=1}$ coefficients.

Interestingly, the process of decrease of $a^{\infty}_{q=1}$ with $p$ does not occur in the same way for $\alpha=0.5$ and for $\alpha=2$. 
For the latter case, the difference between $a^{\infty}_{q=1}$ before and after the resetting is increasing with $p$. Finally, deeply in the sub-volume law phase for participation entropy after the resetting one obtains $a^{\infty}_{q=1}\approx 0$ implying that the state is localized in the Z basis. However, a single cycle of the unitary time evolution noticeably delocalizes the state even at large $p$ resulting in $a^{\infty}_{q=1}$ to be significantly above $0$. In contrast, for the longer-range of interactions ($\alpha=0.5$) the difference between scaling of participation entropy before and after the resetting starts to diminish around $p\approx0.5$, and, in the regime of large $p$, $a^{\infty}_{q=1}\approx 0$ \emph{both} for the wave function before and after the resetting. The differences between the $\alpha=0.5$ and $\alpha=2$ cases trace back to the differences in system size scaling of the entropy of the steady state $\rho_{ss}$. If the system is in the purifying regime then the steady state remains localized in the Z basis throughout the whole cycle of time evolution. In contrast, the mixed dynamics of the system at $\alpha=2$ is able to delocalize the state in the Z basis during a single cycle of time evolution even though the resetting localizes the state entirely.

This subsection shows that the system size scaling of entanglement entropies distinguishes between various dynamical phases of the kicked Ising model with resetting. One of the questions that remain open is whether the wave function becomes fully delocalized in the volume-law phase (and then $D_{q}=\ln 2$ for all $q$). While this seems plausible, the values of the multifractal dimension are smaller than $\ln 2$, and the trend between data for smaller/larger system sizes (dashed vs solid lines in Fig.~\ref{fig:sq}~c),~d)~) appears to suggest that this difference shall not be due to finite volume effects, which would be in line with the conclusions of \cite{Li21}. The long-range order present in the system at $\alpha=0.5$ puts certain constraints on the wave function and prevents full delocalization of the wave function over the whole Hilbert space. Such mechanism is expected to be absent for $\alpha=2$, when there is no long-range order in the system. A careful examination of problems with self-averaging \cite{Solorzano21} and of finite size effects could shed further light on the matter of delocalization of the wave function in the volume-law phase.

\subsection{Eigenstates in the X basis and multifractality}

\begin{figure}
\vspace{0.25cm}
	\includegraphics[width=1\columnwidth]{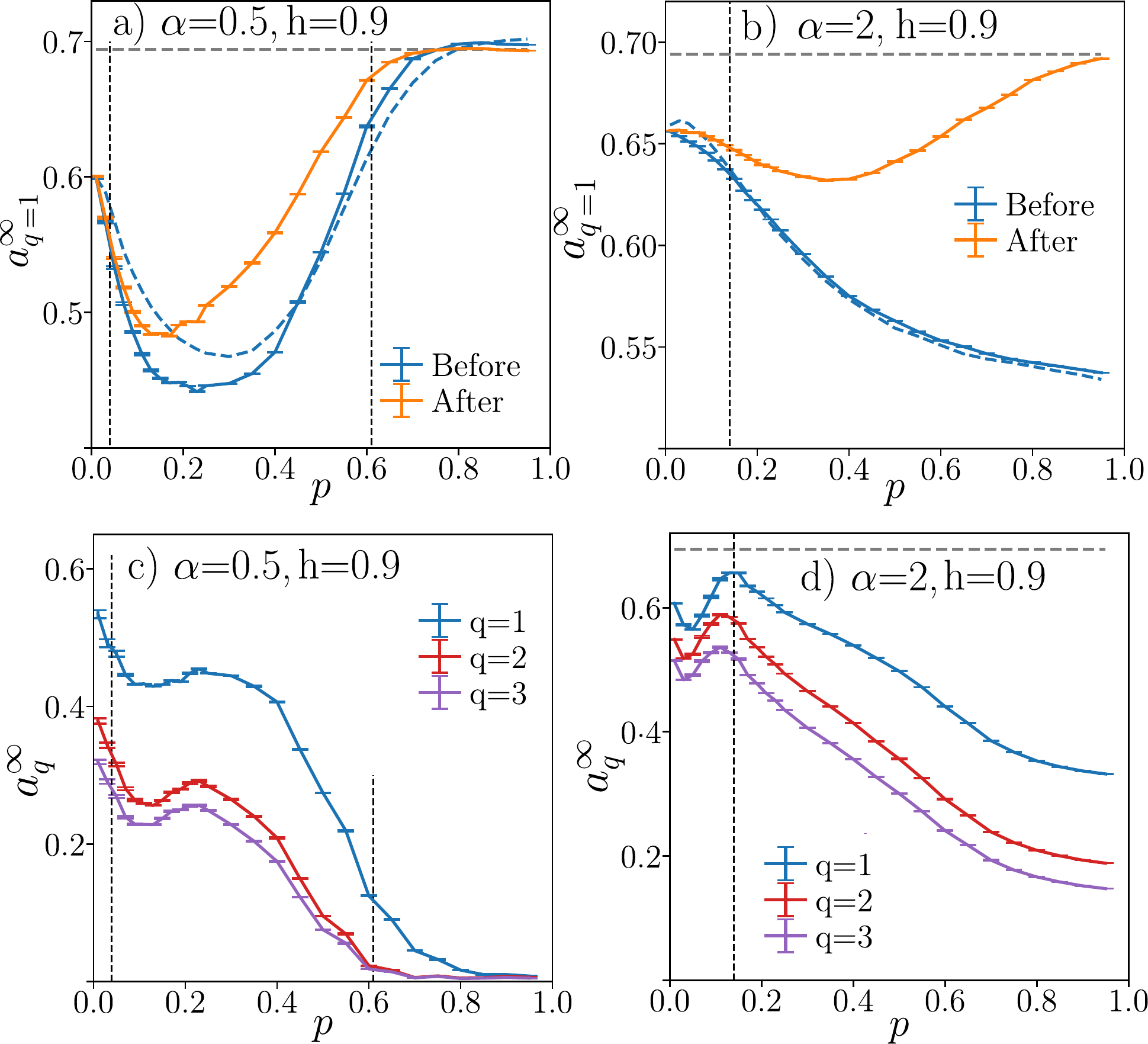}
	\caption{Scaling of participation entropies $S_{q}$ in the steady state of the system. Panels a) and b): the coefficient $a_{q=1}^{\infty}$ governing growth of $S_{q=1}$ in the X basis, respectively for range of interactions $\alpha=0.5$ and $\alpha=2$; the dashed blue lines corresponds to extrapolations based on system sizes $L=8,...,14$, whereas the solid lines correspond to extrapolations based on $L=16,...,24$. Panels c) and d) show a comparison of the coefficients $a_{q}^{\infty}$ for $q=1,2,3$ (derived from $S_{q}$ measured before the resetting), respectively for  $\alpha=0.5$ and $\alpha=2$; the steady state is multifractal when $a_{q}^{\infty}<\ln 2$ and $a_{q}^{\infty}$ are different for different $q$.
  }
	\label{fig:sqApp}
\end{figure}

The participation entropies \eqref{Eq:PartEnt} can be also calculated in the eigenbasis of $\sigma^x_i$ operators (the 'X basis'). This leads to a different picture that is complementary to the previous choice. For instance, the fully localized state in the Z basis $\ket{ \downarrow \downarrow ...\downarrow }$, relevant for the limit of large $p$, becomes an equal superposition of all basis states in the X basis. 

The results for the $a^{\infty}_{q=1}$ coefficient, calculated in the X basis are shown in Fig.~\ref{fig:sqApp}~a),~b). The aforementioned difference with respect to the Z basis is indeed visible. However, the results in the X basis are fully consistent with results in the Z basis: the volume-law phase is characterized by $a^{\infty}_{q=1}$ close to (but smaller than)  the ergodic value $\ln 2$; once the resetting probability increases beyond the critical value $p_c$, the coefficient $a^{\infty}_{q=1}$ drops down and a difference between data before and after the resetting shows up. Finally, the difference becomes small in the purifying steady state regime at large $p$ in the system with $\alpha=0.5$, in contrast to the regime of mixed dynamics for $\alpha=2$. 

Fig.~\ref{fig:sqApp}~c),~d) show the coefficient $a^{\infty}_{q}$ calculated for participation entropies with $q=1,2,3$ in the Z basis using data before the resetting (the results in the X basis are analogous). The coefficients $a^{\infty}_{q}$ are vanishing in the purifying regime at large $p$ in the model with $\alpha=0.5$, showing that the state is indeed localized in the Z basis. For smaller values of $p$ at $\alpha=0.5$, as well as for the entire interval of resetting probabilities for $\alpha=2$, we observe that the values of $a^{\infty}_{q}$ lie between $0$ and $\ln 2$ and are not-equal to each other. This suggests that the steady states in the kicked Ising model with resetting are either fully localized in the Hilbert space or possess a multi-fractal structure for $p>p_c$. The situation is less clear cut in the volume-law phase. For $p<p_c$ the coefficients $a^{\infty}_q$ are quite close to $\ln 2$ for $\alpha=2$ and it is unclear whether the wave function becomes fully delocalized over the Hilbert space in the thermodynamic limit. In contrast, for $p<p_c$ the coefficients $a^{\infty}_q$ are significantly smaller that $\ln 2$ for $\alpha=0.5$ which is a signature of the long-range order in the kicked Ising model with resetting.

\section{Experimental realization with trapped ions}\label{Sec:Exp}

In the above sections we have shown how different types of phase transitions manifest in a spin system evolving according to a Floquet evolution based on long-range interactions and global rotations interspersed with random non-unitary local reset operations. In the following we will outline how to observe all these different types of non-equilibrium phase transitions in a realistic experimental setting based on laser cooled atoms confined in radiofrequency traps.

\subsection{How to realize unitary and non-unitary evolution}\label{sec:ExpA}

\paragraph{Hamiltonian dynamics. - } Trapped-ion systems can directly realize the long-range interacting spin models of the type described in Eq. (\ref{Eq:IsHam}) by off-resonantly coupling pseudo-spin degrees of freedom (detailed below) to the motional collective modes stemming from ion-ion Coulomb interactions \cite{monroe2021programmable,schneider2012experimental}. In particular, given the normal modes frequencies $\omega_m$, the laser Rabi frequency $\Omega$ and the detuning $\mu$ from the center of mass mode, the spin-spin interaction $J_{ij}$ between ions $i$ and $j$ can be explicitly calculated as follows:
\begin{equation}
\label{eq_}
J_{ij}=\Omega^2 \omega_{\rm rec} \sum_{m=1}^L \frac{b_{im}b_{jm}}{\mu^2-\omega_m^2},
\end{equation}
where $\omega_{\rm rec} = \hbar (\Delta k)^2/2M$ is the recoil frequency associated with the transfer of momentum $\hbar (\Delta k)$ and $b_{im}$ is the normal mode transformation matrix that specifies the participation of the $i$-th ion on the $m$-th normal mode. The spin-spin interaction can be approximated with a tunable power law:
\begin{equation}
\label{eq_1}
J_{ij}=\frac{J_0}{|i-j|^\alpha},
\end{equation}
where the power law exponent $\alpha$ can be set by changing the detuning $\mu$ and the trap parameters. Usually the laser detuning is $\mu>\omega_m, \forall m$, giving rise to positive anti-ferromagnetic interactions. However, changing  the relative sign of the transverse field $h$ and the spin-spin interaction $J_0$ gives access to dynamics of both ferromagnetic and anti-ferromagnetic systems \cite{Jurcevic2017,Kaplan2020,tan2021domain,joshi2021observing}. This holds also in the case of the driven-dissipative dynamics under study, as the Kraus operators defined in Eq. (\ref{Eq:rho_meas}) are real.

In order to implement the transverse field Hamiltonian $H_T$ (Eq. \ref{Eq:IsHam}) for the kicked model, a series of global rotations can be applied with the same laser beams generating the Ising Hamiltonian, adding negligible experimental overhead. Another possible strategy is to realize only a single unitary chapter, evolving under the Hamiltonian $H_I+H_T$ by using asymmetric laser detunings $\mu_\pm$ \cite{monroe2021programmable}. This approach would not realize the kicked model, treated here for conceptual simplicity, but all the results derived in this work are not expected to differ qualitatively in the case of unitary evolution under $H_I+H_T$.
This unitary evolution can be easily digitized and expressed in terms of single and two-qubit gates. This makes this model accessible to both analog and digital trapped-ion quantum simulators.

\paragraph{Optical pumping and measurements. -} The other ingredient necessary to experimentally observe the non-equilibrium phase transitions is local measurement and/or resetting operations. 
In this respect, trapped-ion qubits offer high fidelity state detection \cite{noek2013high, Christensen2020high} based on spin dependent fluorescence. Additionally, state preparation can be efficiently realized with optical pumping \cite{happer1972optical}, a well-known technique based on photon scattering and atomic selection rules that effectively implements the operators $K_{i0}$ and $K_{i1}$ leading to the Kraus map defined in Eq. (\ref{Eq:rho_meas}). 

As a non-unitary operation, optical pumping has a fundamental advantage with respect to full detection: it leads to negligible cross-talks, making the reset operation truly local, as it leaves untouched neighboring qubits. As outlined below, the symmetry-breaking transition can be observed by averaging over quantum trajectories and, therefore, does not require projective measurements to postselect the data. On the other hand, MIPTs can be observed only by measuring properties of the quantum state, therefore it is necessary to postselect over quantum trajectories via strong measurements on the randomly selected qubits.

In the following sections, the experimental strategies and challenges to observe both the symmetry-breaking and the entanglement transitions are outlined.

\subsection{Measuring the transitions}
 
\subsubsection{Symmetry breaking transition}

Our results from Section \ref{Sec:order} indicate that ferromagnetic ordering transition of the average state can be observed in the kicked Ising model with resetting: the measurement of the two-point correlation function $X^2$ (see Eq. \eqref{eq:XX}) or the Binder cumulant (see Eq. \ref{Eq:Binder}) allows to pin-point the transition for sufficiently large systems. We stress that, thanks to single atom resolution, both two-body \cite{richerme2014nonlocal,Pagano2020quantum, tan2021domain} and four-body correlators \cite{Islam2013} are routinely measured in trapped-ion systems.
In particular, since these observables are measured averaging over different quantum trajectories, post-selection is not required and resetting via optical pumping is sufficient to probe the transition. 

\begin{figure}[t!]
\vspace{0.25cm}
\centering
\includegraphics[width=1\columnwidth]{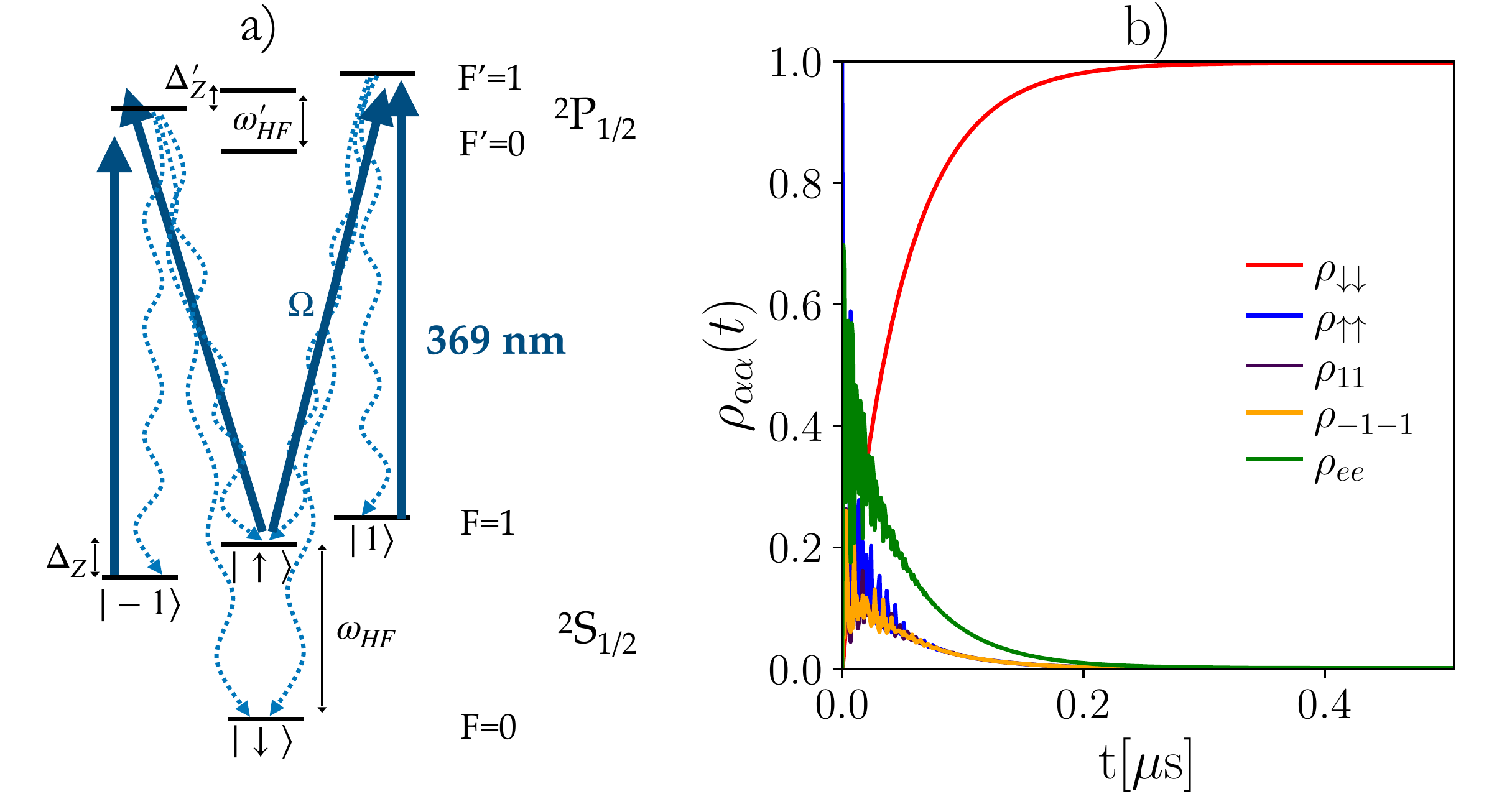}
\caption{
{\bf Optical pumping in $^{171}$Yb$^+$:} a) near-resonant 369 nm light pumps the electron in the dark state $\ket{\downarrow}$, being detuned by $\sim 750 \Gamma$ from the next available transition, with $\Gamma$ being the linewidth of the excited state. $\omega_{HF}=12.6$ GHz and $\omega'_{HF}=2.105$ GHz are the hyperfine splitting of the ground $^2S_{1/2}$ and excited $^2P_{1/2}$ state, respectively. 
{\bf b)}: Master equation simulation of state population evolution with $\Omega=15 \,\Gamma$, ground (excited) state Zeeman splitting $\Delta_Z=0.4\, \Gamma (\Delta'_Z=0.13 \,\Gamma)$ and laser resonant with the $\ket{F=1,m_F=0}\rightarrow \ket{F'= 1,m_F=0}$ transition. Natural linewidth of the $^2P_{1/2}$ state is $\Gamma=2\pi\times19.6$ MHz. The green line is the total of all four excited states of the $^2P_{1/2}$ manifold. We assumed beam waist $w_0=1.5\, \mu$m and power $P= 1\,\mu$W.
}
\label{fig_OP}
\end{figure}

In order to analyze the experimental feasibility, let us consider the concrete example of $^{171}$Yb$^+$ ion with the spin degree of freedom encoded in clock states $\ket{\uparrow}={^2S}_{1/2}\ket{F=1,m_F=0}$ and $\ket{\downarrow}=\ket{F=0,m_F=0}$, respectively (see Fig. \ref{fig_OP}a).
In this case the optical pumping process is carried out by driving the $^2S_{1/2} F=1\rightarrow {^2P}_{1/2} F'=1$ transition at 369 nm. 

The pumping process can be described by a Lindblad Master Equation. A numerical solution of the equation is plotted in Fig. \ref{fig_OP}b which shows that reasonable experimental parameters guarantee $99.8\%$ optical pumping efficiency in $ \sim 500$ ns. By numerically integrating the total excited state population over time, the total average number of scattered photons is predicted to be $\langle N\rangle_{ph}\sim 3$ leading to an average number of photons absorbed by the neighboring ions of $\langle N\rangle_{ph} \sim 1\times10^{-3}$, assuming 4 $\mu$m ion-ion spacing. Therefore, errors in the reset operations are likely to be dominated by optical aberration in the production of a beam array with radius $w_0 <2\,\mu$m, rather than scattering from neighboring ions during pumping. Crosstalk due to stray beam intensity on neighboring ions can be reduced down to $<10^{-4}$ \cite{Shih2021reprogrammable}, leading to an $<0.5\%$ error.

Additionally, another experimental observable related to the average state can be used to detect the onset of the symmetry breaking transition. Recent progress in stochastic sampling of wave function \cite{Brydges19,Leseleuc19,Chiaro20,Pagano2020quantum,Scholl20,Ebadi20,Zeiher17,Veit21} suggests  an alternative route of identification of phase transitions based on detecting changes in structure of a set of wave function snapshots \cite{Mendes-Santos21,Mendes-Santos21b}. In fact, the entanglement phase transition in stabilizer circuits can be seen as a data structure transition \cite{Turkeshi21b}. In that approach, the intrinsic dimension $I_d$ \cite{Facco17} of the data set that encodes the wave function of the stabilizer circuit was demonstrated to be a good probe of the entanglement phase transition. The intrinsic dimension admits a minimum both at the measurement induced phase transition \cite{Turkeshi21b} as well as at the equilibrium phase transitions \cite{Mendes-Santos21b}. 

This motivates us to investigate the following procedure of probing the transitions in the kicked Ising model with resetting. We assume the system to be in the steady state for fixed values of parameters $L$, $p$, $h$ and $\beta$, and take a certain quantum trajectory $\ket{\psi} = \sum_{\beta=1}^{2^L} \psi_{\beta} \ket{\beta}$, where $\ket{\beta}$ are the states of the X basis. We simulate the process of taking snapshots of the wave function by drawing randomly $m$ states from the X basis, in such a way that the probability of drawing state $\ket{\beta}$ is equal to $p_{\beta}^2$. This results in a sequence of $m$ states $(\beta_1,\beta_2,..., \beta_m)$, equivalent to binary number of length $m \times L$, which becomes the first entry of our data set $G$. Generating an independent state $\ket{\psi'}$ (by performing $4L$ cycles of the unitary evolution and resetting, or by restarting the time evolution), we repeat the procedure, obtaining another binary number of length $m \times L$, which becomes the second entry of the data set $G$. The procedure is iterated until the data set consists of $n$ binary numbers. In our calculations we use $m=10$ and $n=5000$. We stress that a database of such dimensions can be readily collected in trapped-ion system, where the result of each single measurement on $L$ qubits is a binary outcome of $L$ bits that gives access to correlators of any order, up to the bitstring probability \cite{Zhang2017observationDPT,Pagano2020quantum}. 

Subsequently, we employ the two nearest-neighbors technique to estimate the intrinsic dimension of the data set $G$. For each element $x$ of $G$ we compute the first and second nearest-neighboring distances $r_1(x)$ and $r_2(x)$ using the xor distance metric, that allows us to obtain the intrinsic dimension $I_d$ of $G$ \cite{Facco17}.  

\begin{figure}
\vspace{0.25cm}
	\includegraphics[width=1\columnwidth]{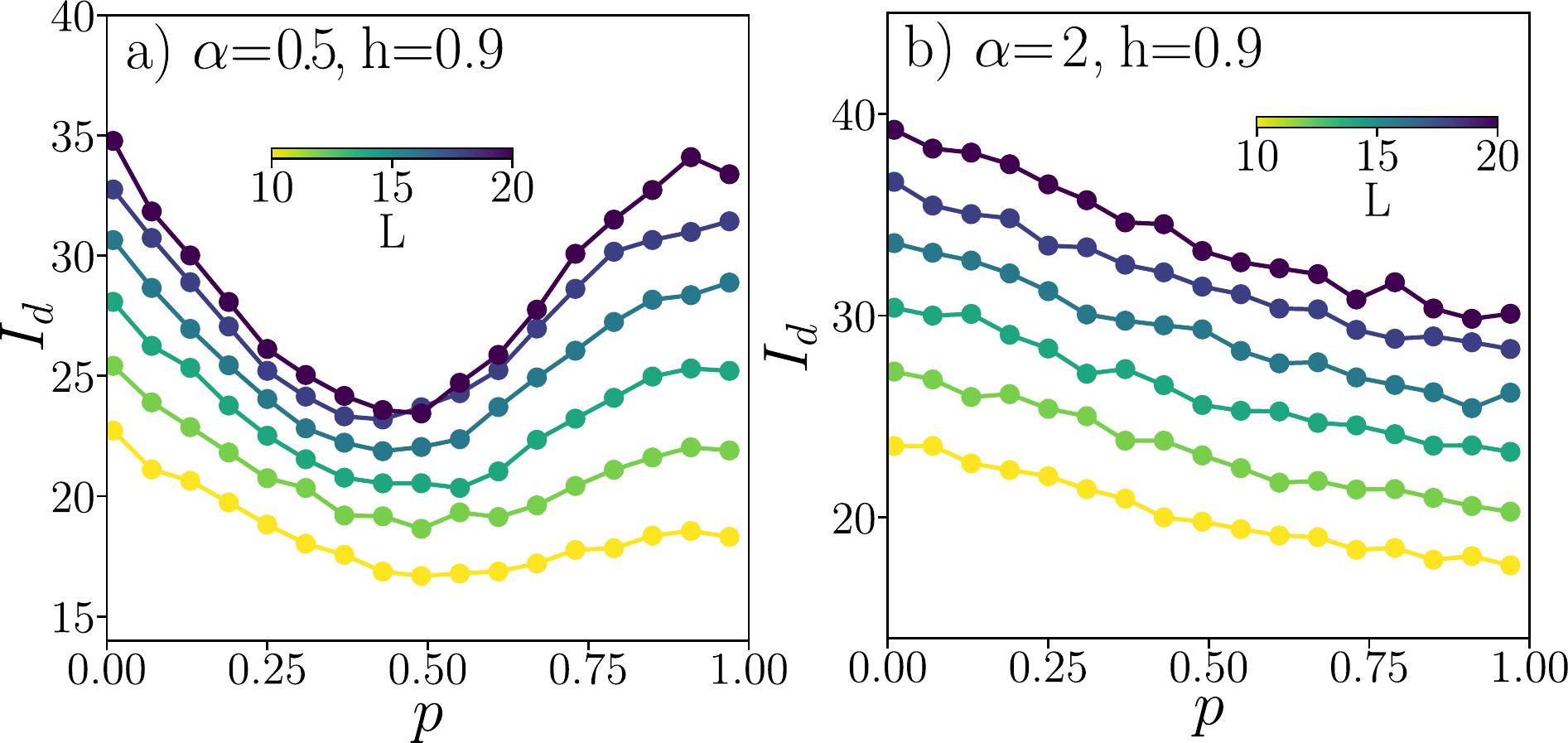}
	\caption{Intrinsic dimension $I_d$ of the set $G$ of wave function snapshots as a function of resetting probability $p$. Panel a) shows the results for system with long-range interactions ($\alpha=0.5$), whereas results for a model with interactions of shorter range are shown in panel b). 
  }
	\label{fig:Id}
\end{figure}

The results for the intrinsic dimension are shown in Fig.~\ref{fig:Id}. We observe that the intrinsic dimension as a function of $p$ has a minimum around $p\approx0.5$  for long-range interactions $\alpha=0.5$ (compatible with $p^{\rm SB}_c=0.58$, see Fig. \ref{fig:bind}) and that there are no distinctive features when $\alpha=2$. This suggests that despite the broad minimum, the computed intrinsic dimension is sensitive to the symmetry breaking phase transition, at least for the system sizes we consider, for which the pin-pointing of the transition with the correlation function is also not straightforward.

On the other hand, the intrinsic dimension remains agnostic to the entanglement transition in our system. This conclusion is not inconsistent with the findings of \cite{Turkeshi21b}, since in that case, the set $G$ encoded full information about the state in the stabilizer circuit. In our case, the set $G$ contains only a fraction of information about the steady state of the system. Such an amount of information is sufficient to find signatures of rapid changes in the properties of the average state of the system but is not sufficient to observe the entanglement phase transition. 

\subsubsection{Entanglement transition}\label{subsec:Expenta}

As explained above, the experimental observation of the entanglement phase transition requires measuring properties of the quantum state of the system such as entanglement entropy. One possible scheme to detect the entanglement phase transition would be to directly measure the bipartite entanglement entropy along individual quantum trajectories in the steady state using quantum interference \cite{Daley12, Abanin12} or random measurements \cite{Enk12,Elben18} and then to repeat the analysis of Sec.\ref{subsec:Expenta}. However, even if feasible in principle, such approaches would require resources exponential in system size $L$ \cite{Huang20}, which makes it a daunting challenge. Moreover, post-selection on the experimental data is necessary to ensure that the average is performed over the trajectories with the same measurement outcomes. This causes an additional overhead in the number of measurements that scales exponentially as $2^{2 p L^2}$ \cite{Bao2020}. However, the post-selection experimental overhead is more bearable if $p_c$ is made as small as possible by decreasing the power-law exponent $\alpha$, as shown in Fig. \ref{fig:qmi}b. Hence, the tunability of the interaction range in the trapped-ion system offers a viable strategy to observe experimentally MIPTs.

In order to perform post-selection, the randomly selected qubits need to be projected in a well defined state via strong measurement prior to resetting. This leads to a more practical experimental challenge: photon scattering that results from spin dependent fluorescence used in the measurement will heavily affect the nearby ions which, in a typical ion chain, are only a few $\mu$m apart. Indeed, in order to differentiate the two spin states, it is necessary to scatter $N_{ph}\sim 10^3$ photons and this leads the nearby ions to absorb on average $N_{ph}\sim1$ photon (considering 4 $\mu$m of ion-ion spacing), hindering the possibility to perform well-controlled local non-unitary operations. A possible solution is qubit hiding: in most ions there are long-lived states  connected to the ground state manifold via narrow optical transitions (e.g. $^2D_{3/2}$ and $^2F_{7/2}$ in Yb$^+$ or $^2D_{5/2}$ in Ba$^+$). The ions' electronic wavefunction can be temporarily stored in those states and not be subject to the measurement in order to avoid crosstalks. The fundamental limitation of this approach is given by the probability that one ion decays to the ground state before the measurement is over, namely $e^{-L t_{\rm meas}/\tau}$, where $ \tau$ is the metastable state lifetime and $t_{\rm meas}$ is the measurement time, which can be pushed to a few tens of $\mu$s with high resolution optics \cite{Noek13}. Considering typical metastable state lifetimes (e.g. $\tau\sim 60 \,{\rm ms}$ for $^2D_{3/2}$ in Yb$^+$\cite{Schacht2015Yb}, $\tau\sim 26$ s for $^2D_{5/2}$ in Ba$^+$ \cite{Mohanty2015lifetime} or $\tau\sim3700$ days for $^2F_{7/2}$ in Yb$^+$ \cite{Roberts1997Observation,yang2021realizing}), this error can be made negligible depending on the atomic levels in use. Qubit hiding also allows to use the ions that have been measured for sympathetic recooling to eliminate the heating of the shared motional modes induced by the measurement. Since the hidden qubits are encoded in a different electronic state, they are insensitive to cooling lasers, which is also a crucial requirement to retain high-fidelity in quantum error correction applications \cite{Erhard2021entangling}.

An alternative route to detect the entanglement transition proposed in \cite{Gullans2020} gets rid of the exponential requirement related to the quantum state observable but still requires an exponential number of measurement to average over different quantum trajectories. In this scheme, the spin with number $i=1$ is treated as a \emph{reference qubit}, whereas the spins $i=2,..., L$ are treated as \emph{the system}. Initially, the reference qubit is strongly entangled with the system, so that the initial state reads
\begin{equation}
\label{Eq:probe}
\ket{\psi}= \frac{1}{\sqrt{2}} \left( \ket{\uparrow}_R\ket{\phi_1}_S + \ket{\downarrow}_R\ket{\phi_2}_S \right),
\end{equation}
where $\ket{\phi_{1,2}}_S$ are (nearly orthogonal) states of the system (spins $i=2,..., L$). To prepare such a state, we perform $2L$ cycles of the unitary evolution \eqref{Eq:IsHam} with resetting of all of the spins with probability $p=0.01$. This produces a strong entanglement of the reference qubit with the rest of the system. Alternatively one could entangle the reference qubit with the system applying a series of $L-1$ consecutive XX gates \cite{debnath2016demonstration} (see appendix \ref{sec:XXpreparation} for details).
Subsequently, we set $t=0$ and start the evolution of the system with the the unitary evolution \eqref{Eq:IsHam} restricted only to spins of the system ($i=2,...,L$) and performing the resetting of the spins of the system with probability $p$. The idea is that in a volume-law phase the measurements do not collapse the state of the system and hence the reference qubit remains strongly entangled with the system. In contrast, when resetting collapses the state of  the system, the state of the reference qubit is purified. This can be diagnosed by looking at the von Neumann entropy $S_R$ of the reduced density matrix $\rho_R$ of the reference qubit
\begin{equation}\label{Eq:RefQubS}
S_R=-\text{Tr}_R(\rho_R\ln\rho_R);\qquad\rho_R=\text{Tr}_{S}(\ket{\psi}\bra{\psi}),
\end{equation}
where $R$ denotes the reference qubit and $S$ denotes the system (spins $i=2,...,L$). The data requirements in this case are more lenient as the entanglement entropy of a single qubit requires only measurements in three orthogonal basis.

\begin{figure}[!t]
\vspace{0.25cm}
	\includegraphics[width=1\columnwidth]{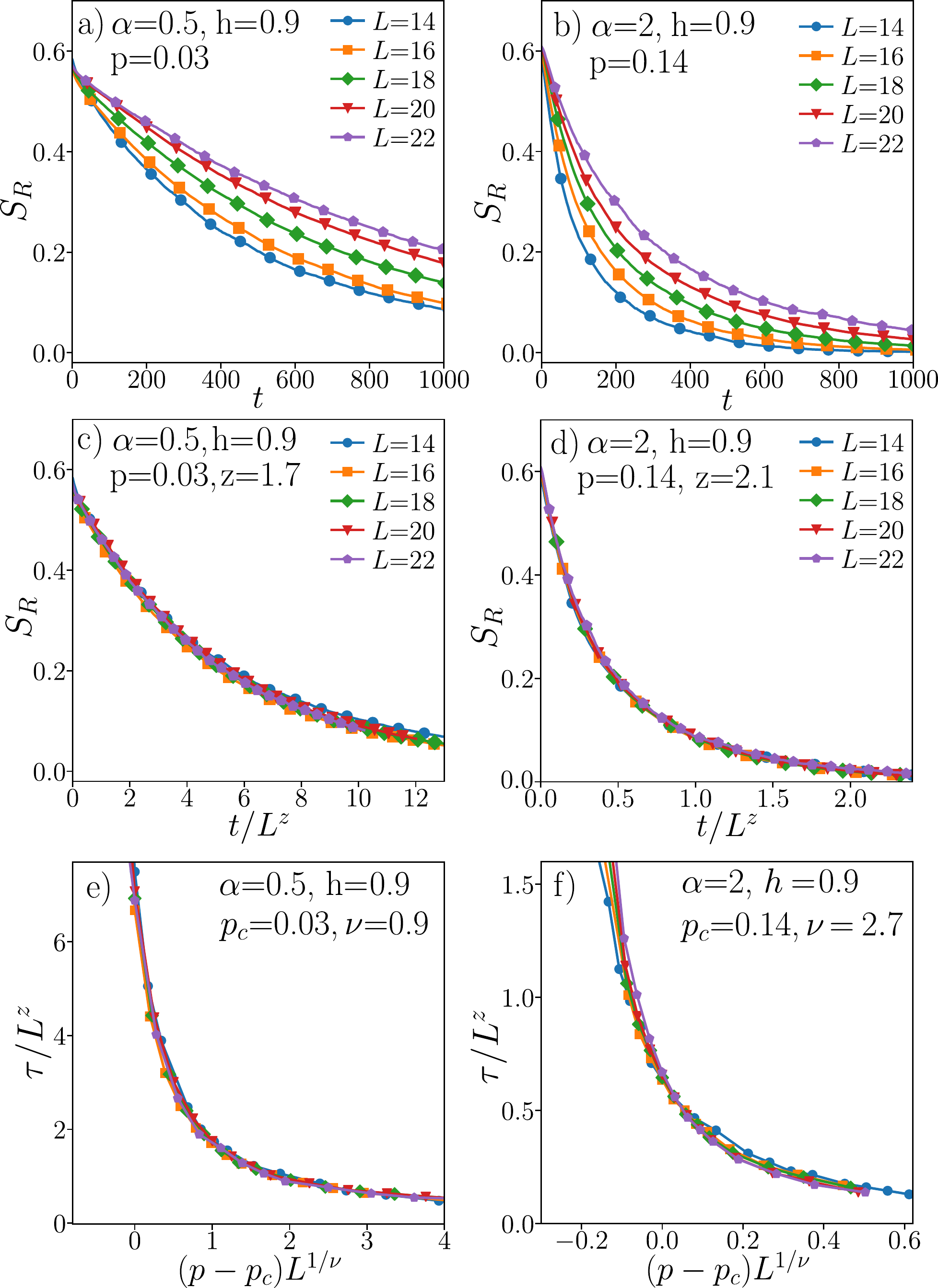}
	\caption{Probing the entanglement phase transition with a single reference qubit. Panels a) and b) show the time evolution of entanglement entropy $S_R$ of the reference qubit (measured before the resetting) respectively for $\alpha=0.5$ and $\alpha=2$, values of the resetting probability are close to the estimated position of entanglement phase transition. The same is shown in Panels c) and d) but the time is now rescaled by a factor $L^z$ which leads to a collapse of the data. Panels e) and f): time $\tau$ such that $S_R(\tau)=s_0=0.15$, rescaled by $L^z$ is shown as a function of $(p-p_c)L^{1/nu}$. 
  }
	\label{fig:probe}
\end{figure}

The results are shown in  Fig.~\ref{fig:probe}~a),~b) for values of resetting probability close to the estimated entanglement transition. We observe that the decay of $S_R$ in time becomes gradually slower with increasing system size, both for $\alpha=0.5$ and $\alpha=2$. The emergent conformal symmetry at the entanglement phase transition in stabilizer circuits implies that for $p=p_c$, $S_R$ becomes a universal function of $t/L$. We find a collapse of $S_R$ at $p \approx p_c$ (with $p_c$ found in Sec.~\ref{Sec:entanglement}) as a function of $t/L^z$, see Fig.~\ref{fig:probe}~c),~d) with $z \neq 1$. This is another manifestation of the lack of conformal invariance at the entanglement transition in our model.  The dynamical exponent $z$ is larger than the value consistent with conformal field theory $z_{CFT}=1$. In contrast, the dynamical exponents found for stabilizer circuits with long-range interactions are smaller than unity \cite{Block2021}.

The entanglement entropy of the reference qubit follows similar scaling laws as the QMI discussed in Sec.~\ref{Sec:entanglement}. To demonstrate this, we introduce a time $\tau$ such that $S_R(\tau)=s_0=0.15$ (other values of $s_0 \in[0.1,0.4]$ yield similar results) and perform a finite size scaling analysis of the time $\tau$ using the following ansatz 
\begin{equation}\label{eq:scalingProbe}
    \tau=L^z g[ (p-p_c)L^{1/\nu}],
\end{equation}
 where $g(y)$ is an universal function, the value of $z$ is fixed from the collapse of $S_R$ at the transition point and $p_c$, $\nu$ are treated as free parameters. The collapses are shown in Fig.~\ref{fig:probe}~e),~f). The obtained values of the critical resetting probability $p_c$ and of the exponents $\nu$ are consistent with the results obtained for QMI in Sec.~\ref{Sec:entanglement}, showing that the entanglement entropy $S_R$ of the reference qubit indeed probes the critical behavior at the entanglement phase transition in the system. However, the predictive power of the method investigated in this section is limited. To extract the value of the dynamical critical exponent $z$ we had to know the value of the critical resetting probability, $p_c$, in advance. Finding the collapse of the entanglement entropy $S_R$ of the reference qubit as function of $t/L^z$ for $p\approx p_c$, we determined the value of $z$. However, collapses of similar quality can be obtained for a resetting probability $\tilde p$ that differs from $p_c$ by a factor of $50\%$, leading to a spurious value of the dynamical critical exponent $\tilde z$ that is significantly different than the value $z$ obtained in the collapse for $p\approx p_c$. Performing the collapse of $\tau/L^{z}$ leads to a spurious value of the critical resetting probability, self-consistently close to the value $\tilde p$ for which $\tilde z$ was determined.
 
Summing up, the entanglement entropy of the reference qubit can be used to find the value of the dynamical critical exponent $z$ for a generic entanglement phase transition if the position of the transition is known in advance. If there is an emergent conformal symmetry at the transition, as for instance in stabilizer circuits, the value of $z$ is fixed to $1$, which can be used to pin-point the position of the transition. This is no longer the case in the absence of conformal symmetry, as we have shown for the kicked Ising model with resetting. In such case, the investigation of the entanglement entropy of the reference qubit allows for a measurement of the dynamical exponent $z$ given the information about $p_c$ that can be either extracted from numerical analysis of the QMI or measured.  

\subsubsection{Effect of noise}\label{subsec:noise}

Here we briefly discuss the effects that the presence of noise (interaction with the environment, errors in the experimental protocol, etc.) has on both the DPT and the MIPT.

For what concerns the symmetry breaking transition, the average steady state is already mixed, so that coupling to some additional small noise will simply shift the critical probability $p_c$ in favor of the paramagnetic region. For example, this behavior can be observed more quantitatively by introducing a small random noise in the magnetic field $h+h_{\xi}$ and then average over the distribution $\mathcal{P}(h_{\xi})$ of the noise. This type of noise is one of the main sources of decoherence in a trapped-ion system. In the simple case of a Gaussian probability distribution we find
\begin{equation}\label{Eq:noise_h}
\langle p_c\rangle_{\xi}=\int d\xi \mathcal{P}(h_{\xi})p_c(h+h_{\xi})\approx p_c+ p_c''(h)\langle h_{\xi}^2\rangle_{\xi}/2
\end{equation}

\begin{figure}[!t]
    \centering
    \includegraphics[width=\columnwidth]{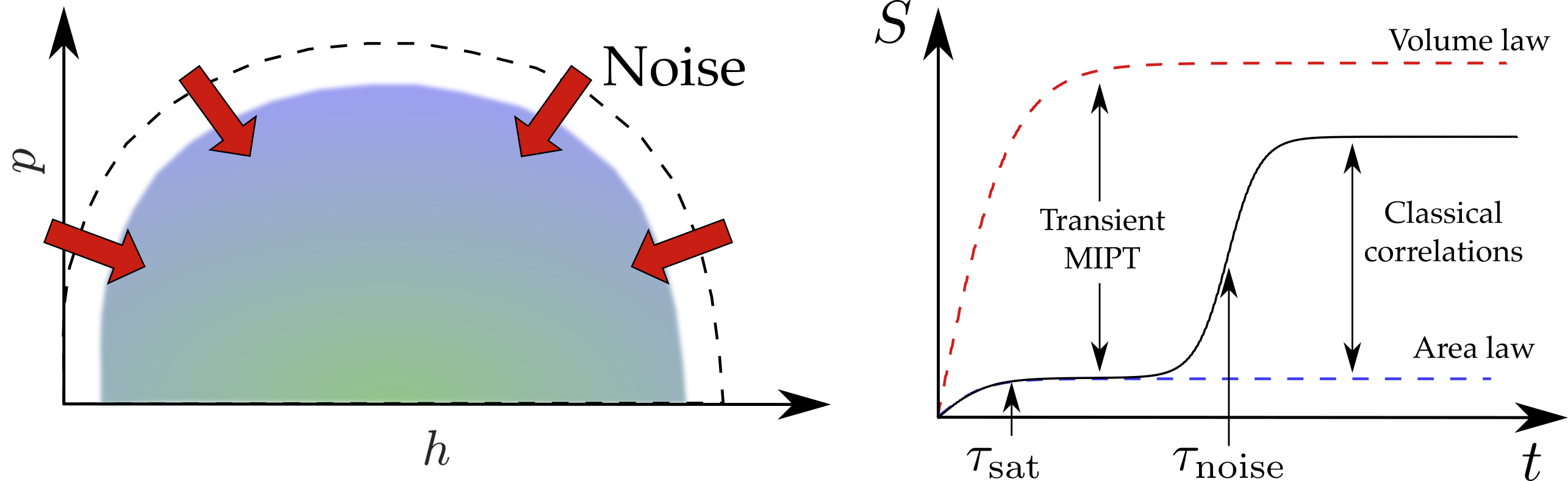}
    \caption{Sketch of the effect of noise on DPT (left) and MIPT (right). Additional noise results in larger mixedness of the system and shift the critical probability in favor of the paramagnetic region (red arrows). For what concerns the MIPT, coupling to the environment leads to the build up of classical correlations over a timescale $\tau_{\rm noise}$; if the saturation timescale $\tau_{\rm sat}$ for the entanglement entropy is smaller than $\tau_{\rm noise}$, the MIPT is still visible as a metastable phase transition.}
    \label{Fig:noise}
\end{figure}

The small corrections to the critical probability are proportional to the second derivative of $p_c(h)$ and thus always negative, see Eq. \eqref{Eq:MFlinpc} and Fig. \ref{fig:Color_map1} and \ref{Fig:noise}.

On the other hand the effect of environment noise on the entanglement transition is more severe. In particular, the sub-volume phase of the entanglement entropy is metastable and disappears at long times, when the environment dynamics becomes predominant. In this regime, the entropy stops being a good measure of the entanglement since the system builds up classical correlations due to the coupling to the environment. Nonetheless, the measurement induced transition in the entanglement may still be detected in the transient dynamics of the system if the crossover time to the volume phase is long enough. This crossover time $\tau_{\textrm{noise}}$ depends on the details of the model and on the strength of the noise, but if $\tau_{\textrm{sat}}$ is the typical saturation time needed to achieve a steady state of the unitary and measurement dynamics, then the metastable non-volume phase is still observable when $\tau_{\textrm{sat}}\ll\tau_{\textrm{noise}}$. In particular our numerical simulations show that $\tau_{\rm sat}\sim L$. Since in trapped-ion systems the time durations  of detection, resetting and single qubit rotations (1-100 $\mu$s) are usually one order of magnitude faster than interaction timescales ($\sim 1$ ms), we can infer that the coherence time needed to observe MIPTs is $\tau_{\rm sat}\sim L/J_0$, with $J_0$ being the NN spin-spin interactions.

\section{Conclusions}\label{Sec:conc}

In this work we proposed a new framework to simultaneously investigate dissipative and measurement-induced phase transitions in experimentally realizable quantum-many body systems, whose dynamics is captured by a dissipative Floquet theory. We considered a kicked Ising model with resetting, and found that both a DPT and a MIPT emerge as a result of the interplay between the unitary dynamics and the random quantum measurements.

The DPT occurs between a ferromagnetic ordered phase and a disordered phase. It can be studied through several observables, such as the magnetization, spin correlation functions or the Binder cumulant, that act as order parameters for the transition and that are \emph{linear} functions of the average state of the system $\rho_{ss}(t)$.

The MIPT in the entanglement entropy has a similar phenomenology to those occurring in hybrid quantum circuits: the bipartite entanglement entropy scales linearly with system size on one side of the transition or saturates on the other side, and the quantum mutual information has a peak at the transition. The MIPT in our model exhibits no conformal symmetry, in contrast to the case of Clifford and Haar circuits \cite{Zabalo20}, but analogous to the MIPT observed in Clifford circuits with sufficiently long-range power-law interactions \cite{Block2021}. The observables employed to study the MIPT are non-linear functions of the quantum state along a particular trajectory, marking a crucial difference with the DPT observables.

In fact, even though the observables used to identify the DPT can be measured on individual quantum trajectories, other information about the measurement induced dynamics is lost in the process of the state average. As a consequence, the average state of the system $\rho_{ss}(t)$, remains agnostic to the underlying transitions at the level of quantum trajectories. Indeed, our findings indicate that the MIPT and the DPT exhibit a different dependence on the range of interactions and occur at significantly different measurement probabilities.   

Nonetheless, the quantum measurements play a qualitatively similar role both for the DPT and the MIPT: they suppress the ferromagnetic long-range order in the system by locally projecting onto a paramagnetic disordered state, and also reduce the amount of entanglement in the state, so that for large resetting probabilities only a disordered sub-volume law phase survives.

The symmetry breaking transition crucially relies on the presence of long-range interactions and ceases to exist once the range of interactions becomes sufficiently short, i.e. $\alpha>\alpha_c\sim1.3$. This behavior is consistent with equilibrium quantum magnets at finite temperature, where sufficiently long-range interactions provide a way around the Mermin-Wagner-Hohenberg theorem~\cite{Mermin66, Hohenberg67} and enable ferromagnetic order at non-zero temperature even in 1D~\cite{Dyson69}. For $\alpha<\alpha_c$, the system develops a long-range order in the steady state if the resetting probability $p$ is smaller than a critical value $p^{SB}_c$. The value of $p^{SB}_c$ vanishes for $\alpha>\alpha_c$ and increases when the range of interactions is increased, reaching its maximum for infinite range interactions, i.e. for $\alpha=0$. The permutation symmetry of our system at the level of super-operators for $\alpha=0$ enabled us to numerically investigate systems of up to few hundred sites: we firmly established the existence and the critical properties of the symmetry breaking transition in our model, showing that they connect smoothly to $\alpha>0$, as indicated by our numerical results. 

On the other hand, the range of interactions plays an opposite role for MIPT. The critical resetting probability $p_c$ for MIPT increases with $\alpha$, making the volume law phase more robust for short range interactions. We have analyzed this trend up to $\alpha=4$, but it is known that MIPTs occur \cite{Li2019} even in spin chains with local interactions ($\alpha\rightarrow \infty$). Conversely, for $\alpha \rightarrow 0$, our analysis indicates that the critical resetting probability for MIPT $p_c\rightarrow 0$ and there is no volume-law phase in our system; due to the limited system sizes accessible to our calculations, we cannot determine whether this a genuine area law phase or simply a phase with sub-volume scaling of entanglement entropy.

Within the model we investigate, ferromagnetic long-range order arises both in the volume law and sub-volume law phases, i.e. there are values of $\alpha$ for which both $p_c$ and $p_c^{SB}$ are non-zero: for sufficiently long-range interactions ($\alpha<\alpha_c\approx1.3$) the system is in a volume-law phase that supports long-range ferromagnetic order, whereas for $\alpha>\alpha_c$ the volume-law phase is disordered. The disordered phase may also coexist with either one of volume law or sub-volume law phase, so that the phase diagram in the $\alpha$-$p$ space is roughly divided into four regions. This observation proves that ferromagnetic order and volume law scaling of the entanglement are not strictly correlated, and highlights the potential of dissipative Floquet theories in investigating the generic features of open quantum systems dynamics. 

To further characterize the properties of the average steady state $\rho_{ss}(t)$, we investigated its purity. Within a mean field approximation, the ordered phase corresponds to a phase in which the state of the system remains mixed, whereas the steady state of the disordered phase is purified. However, our numerical simulations of the average state of the system indicate that for $\alpha>0$, the purity and the ferromagnetic state behave differently: for the DPT the critical resetting probability $p_c$ decreases with $\alpha$, while the extension of the mixed state increases when the range of interactions decreases. The changes in the purity of the average state are a distinct phenomenon from the measurement induced purification transitions \cite{Gullans2020, Choi20} as discussed in Appendix \ref{sec:app_puri}.

In order to provide an alternative perspective on MIPTs we considered the participation entropies of the wave function on quantum trajectories, which allow to quantify what portion of the many-body Hilbert space is covered by the wave function in the volume and sub-volume law phases. Our results indicate that the wave function is nearly fully delocalized over the Hilbert space in the volume-law phase, but its extension decreases when ferromagnetic order is also present. This is intuitively clear: the order parameter may be non-zero only if the wave function has larger coefficients on the basis states corresponding to the ferromagnetic order. In addition, the participation entropies display a multifractal behavior in a wide regime of resetting probabilities beyond the MIPT, and a study of their scaling with system size calculated before and after the resetting process allows to pinpoint the regimes in which the average state is pure.

One key advantage of the new platform we propose is that it is readily realizable in a trapped ions experiments using hardware and techniques that are currently available.  We have shown that the considered Hamiltonian dynamics of spin systems subject to power-law interactions can be implemented by off-resonantly coupling pseudo-spin degrees of freedom to the motional collective modes stemming from ion-ion Coulomb interactions, while the non-unitary resetting operations can be implemented by utilizing a scheme based on optical pumping that we analyze in detail for $^{171}\text{Yb}^+$ ions. While the resources needed to experimentally observe the MIPT scale exponentially with the system size, making it a difficult task, we argued that the DPT can be experimentally observed by direct measurements of the two-point correlation functions or of the Binder cumulants, and have also shown that its position can be located by an analysis of the intrinsic dimension of a set of wave function snapshots. These findings open up many concrete possibilities for the experimental observation of DPTs and MIPTs, which has been lagging behind compared to the theoretical work done on this subject.

Our results prove that the framework of a dissipative Floquet dynamics is a powerful tool to study both DPTs and MIPTs at the same time. It provides a simple yet effective platform to investigate the interplay of these related phenomena and to bridge a gap in our understanding of the consequences of a competition between unitary and dissipative dynamics. Our findings also imply that the phenomenon of measurement induced phase transition may be observed in a much broader class of systems than what has been considered so far in the literature.

The framework of dissipative Floquet evolution can find further applications in the context of both steady state and trajectory transitions and phases. In terms of classes of dynamics, one immediate extensions is to consider models in which only one between dissipation and coherent dynamics is periodic in time, while the other is constant: these situations are also compatible with the dissipative Floquet framework, and only partly explored~\cite{Schnell20}. Moreover, the Floquet formalism easily allows to address external periodic drives (such as for example radiation acting on the system) and may help in drawing a connection to solid state systems, where experiments utilizing periodic driving or pump-probe schemes are increasingly common and easy to realize \cite{Averitt02,Basov11,Morrison14,Mitrano16,Kogar20}. Finally, it would be interesting to frame topological effects in this context \cite{Lavasani21,Bravyi06}. This may allow to find possible contact points between the pure and mixed state description of topological states, and to understand the role of quantum correlations in the unexplored case of density matrices, leveraging on the broad understanding in the context of pure states.  

{\it Note added:} While finalizing this manuscript we became aware of related works \cite{Czischek21, Noel21}.

\section{Acknowledgments}
GP acknowledges illuminating discussions with William Morong, Patrick Becker and Christopher Monroe. PS acknowledges illuminating discussions with Antonello Scardicchio. FS and GC acknowledge useful discussions with Alessio Lerose. The work of GC, FS, XT and MD is partly supported by the ERC under grant number 758329 (AGEnTh), by the MIUR Programme FARE (MEPH), and by the European Union's Horizon 2020 research and innovation programme under grant agreement No 817482 (Pasquans). PS acknowledges the support of  Foundation  for
Polish   Science   (FNP)   through   scholarship START and support by PL-Grid Infrastructure. GP was supported by the NSF CAREER Award (Award No. PHY-2144910), the DOE Office of Science, Office of Nuclear Physics, under Award no. DE-SC0021143, the Army Research Office (W911NF21P0003) and the Office of Naval Research (N00014-20-1-2695).

\appendix
\section{Exact numerics in the permutationally symmetric subspace}
\label{sec:app_perm}

In this section we report additional details on the method used to simulate the dynamics for $\alpha=0$ in the permutationally symmetric subspace.

The key steps are: i) choosing an appropriate basis for the subspace, ii) computing the matrix elements of the observable in the basis, iii) writing the initial state as a vector in this basis, iv) finding a prescription for computing observables.

Step i) was illustrated in the main text, with the definition of the states $\ket{\{n_i\}}$ in Eq.~(\ref{eq:perm_basis}). The normalization constant is $\mathcal{N}_{\{n_i\}}=\sqrt{L! \;n_1!\; n_2!\; n_3!\; n_4!}$.

For step ii), we define the following superoperators:
\begin{equation}\label{Eq:AppXlXr}
    X_l=\sum_i \left(\sigma_i^x \otimes \mathbf{1}_i\right)\qquad X_r=\sum_i\left(\mathbf{1}_i \otimes  \sigma_i^x\right)
\end{equation}
\begin{equation}\label{Eq:AppZlZr}
    Z_l=\sum_i \left(\sigma_i^z \otimes \mathbf{1}_i\right)\qquad Z_r=\sum_i\left(\mathbf{1}_i \otimes  \sigma_i^z\right)
\end{equation}
\begin{equation}\label{Eq:AppO}
    O= \sum_i \Big( \ket{\downarrow\downarrow}\bra{\downarrow\downarrow}+\ket{\downarrow\downarrow}\bra{\uparrow\uparrow}-\mathbf{1}\otimes \mathbf{1}\Big)_i
\end{equation}

The Hamiltonian part of the evolution of $\ket{\rho}\rangle$ can then be written as
\begin{equation}
e^{-iH_TT/2}\rho e^{iH_TT/2} \Longrightarrow \exp\left[- i h(Z_l-Z_r)\frac{T}{2}\right]\ket{\rho}\rangle,
\end{equation}
\begin{equation}
e^{-iH_IT/2}\rho e^{iH_IT/2}\; \Longrightarrow \exp\left[i J(X_l^2-X_r^2)\frac{T}{2}\right]\ket{\rho}\rangle,
\end{equation}
while the measurement process is
\begin{equation}
    \sum_\mu K_\mu \rho K_\mu^\dagger\;\Longrightarrow \exp\left[-\ln(1-p)O\right]\ket{\rho}\rangle.
\end{equation}

The action of the superoperators can be easily written in the basis $\ket{\{n_i\}}=\ket{n1,n2,n3,n4}$, obtaining
\begin{equation}
\begin{split}
Z_l \ket{\{n_i\}}&=(n_1+n_2-n_3-n_4)\ket{\{n_i\}},  \\ 
Z_r \ket{\{n_i\}}&=(n_1-n_2+n_3-n_4)\ket{n_1,n_2,n_3,n_4},  \\ 
\end{split}
\end{equation}
\begin{equation}
\begin{split}
X_l \ket{\{n_i\}}&=\sqrt{n_1(n_3+1)}\ket{n_1-1,n_2,n_3+1,n_4}  \\ 
&+\sqrt{n_2(n_4+1)}\ket{n_1,n_2-1,n_3,n_4+1}\\
&+\sqrt{n_3(n_1+1)}\ket{n_1+1,n_2,n_3-1,n_4}\\
&+\sqrt{n_4(n_2+1)}\ket{n_1,n_2+1,n_3,n_4-1},\\
\end{split}
\end{equation}
\begin{equation}
\begin{split}
X_r \ket{\{n_i\}}&=\sqrt{n_1(n_2+1)}\ket{n_1-1,n_2+1,n_3,n_4}  \\ 
&+\sqrt{n_2(n_1+1)}\ket{n_1+1,n_2-1,n_3,n_4}\\
&+\sqrt{n_3(n_4+1)}\ket{n_1,n_2,n_3-1,n_4+1}\\
&+\sqrt{n_4(n_3+1)}\ket{n_1,n_2,n_3+1,n_4-1},
\end{split}
\end{equation}
\begin{equation}
\begin{split}
    O\ket{\{n_i\}}&=(n_4-L)\ket{n_1,n_2,n_3,n_4}\\
    &+\sqrt{n_1(n_4+1)}\ket{n_1-1,n_2,n_3,n_4+1}.
\end{split}
\end{equation}

For step iii), we now write the initial state in the basis $\ket{\{n_i\}}$. We consider the case of a pure initial state polarized along $+x$. We get
\begin{equation}
\begin{split}
    \ket{\rho}\rangle_{t=0}&=\frac{1}{2^L}\prod_{i=1}^L \big(\ket{\uparrow\uparrow}+\ket{\uparrow\downarrow}+\ket{\downarrow\uparrow}+\ket{\downarrow\downarrow}\big)\\
    &=\frac{1}{2^L}\sum_{\{n_i\}}\sqrt{\frac{N!}{n_1!\,n_2!\,n_3!\,n_4!}}\ket{\{n_i\}}.
\end{split}
\end{equation}

Lastly, we discuss step iv). We note that the scalar product of two vectors $\langle\braket{\rho|\sigma}\rangle$ corresponds to the quantity $\mathrm{Tr}[\rho^\dagger \sigma]$. Therefore, by defining the vector $\ket{\mathbf{1}}\rangle$ associated with the identity matrix, we can compute the expectation value of an observable $A$ on the state $\rho$ as $\braket{A}=\langle\braket{\mathbf{1}|A\otimes 1|\rho}\rangle=\langle\braket{\mathbf{1}|1\otimes A|\rho}\rangle$. The identity vector $\ket{1}\rangle$ can be written in the basis $\ket{\{n_i\}}$ as
\begin{equation}
\begin{split}
    \ket{1}\rangle &= \prod_i \big(\ket{\uparrow\uparrow}+\ket{\downarrow\downarrow}\big)_i\\
    &=\sum_{n_1=0}^L \sqrt{\binom{L}{n_1}}\ket{n_1,0,0,N-n_1}.
\end{split}
\end{equation}

\section{Entanglement phase transition with deterministic state preparation}\label{sec:XXpreparation}
\begin{figure}[!t]
	\includegraphics[width=1\columnwidth]{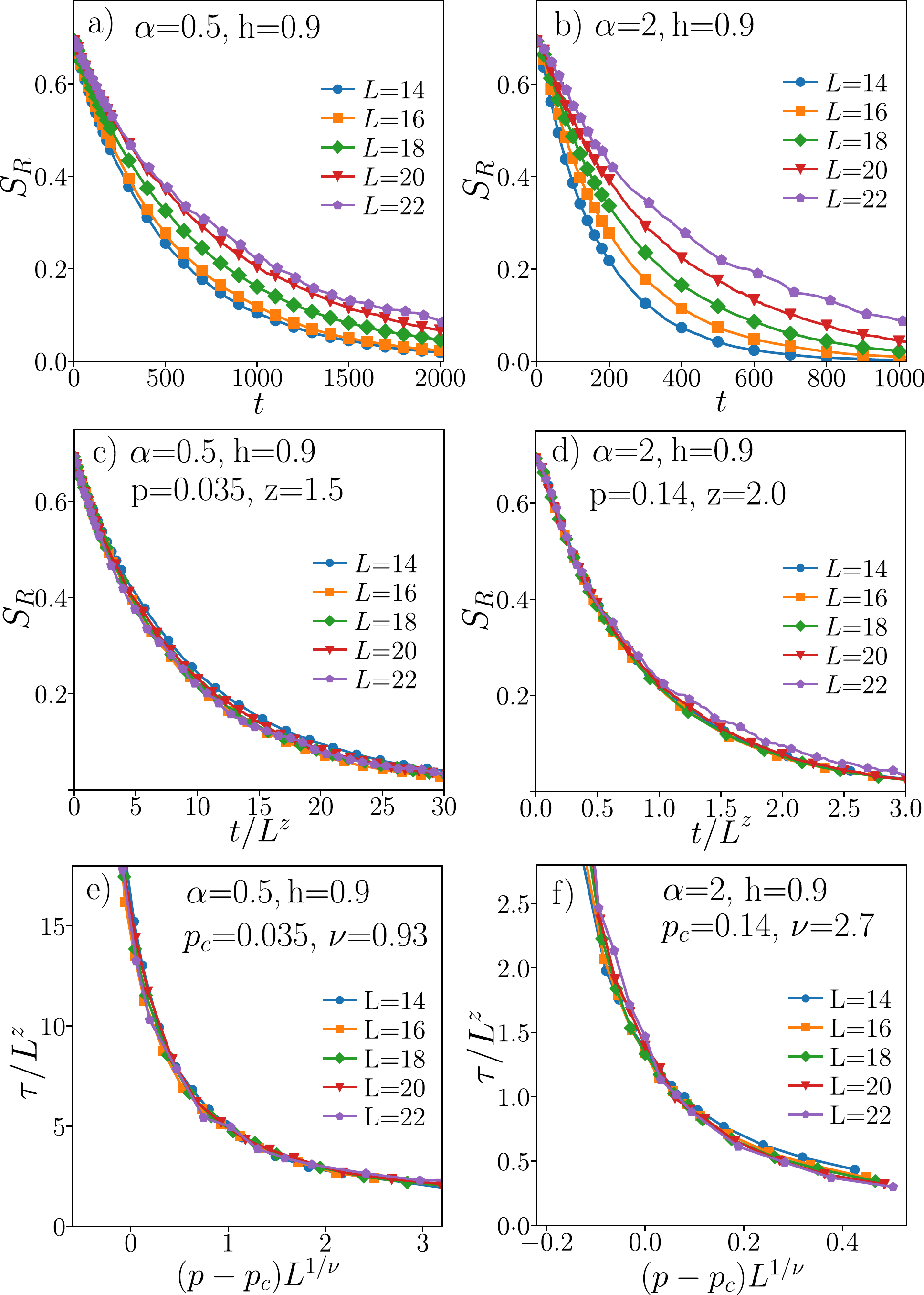}
	\caption{Probing the entanglement phase transition with a single reference qubit. Panels a) and b) show the time evolution of entanglement entropy $S_R$ of the reference qubit (measured before the resetting) respectively for $\alpha=0.5$ and $\alpha=2$, values of the resetting probability are close to the estimated position of entanglement phase transition. The same is shown in Panels c) and d) but the time is now rescaled by a factor $L^z$ which leads to a collapse of the data. Panels e) and f): time $\tau$ such that $S_R(\tau)=s_0=0.15$, rescaled  by $L^z$ is shown as a function of $(p-p_c)L^{1/\nu}$.
  }
	\label{fig:probe_det}
\end{figure}

Here we show that the entanglement phase transition can be measured by entangling sequentially the reference qubit with all the remaining $L-1$ system qubits. The initial state is prepared as follows: we start with a fully polarized state $\ket{\psi}=\ket{\downarrow \downarrow \ldots \downarrow}$, subsequently we use the XX gates: $XX_i=\exp(i \frac{\pi}{4} \sigma^x_1 \sigma^x_i)$, to create the state $\prod_{i=2}^L XX_i \ket{\psi}$ which is our initial state (qubit $i=1$ is the reference, qubits $i=2,...,L$ constitute the system). The results are shown in Fig. \ref{fig:probe_det} and are in agreement with the probabilistic state preparation of a volume law state shown in Fig. \ref{fig:probe}.

\bigbreak
\newpage

%

\end{document}